\documentclass[useAMS,usenatbib]{mn2e}

\usepackage{amsmath,array,amsfonts}
\usepackage{amssymb}
\usepackage{graphicx}
\usepackage{natbib}
\usepackage{pstricks}
\usepackage{verbatim}
\usepackage{multirow}
\usepackage{bbold}
\usepackage{pifont}

\newcommand{\ltsima} {$\; \buildrel < \over \sim \;$}
\newcommand{\gtsima} {$\; \buildrel > \over \sim \;$}
\newcommand{\lta} {\lower.5ex\hbox{\ltsima}}
\newcommand{\gta} {\lower.5ex\hbox{\gtsima}}

\def\smica{{\tt SMICA}}
\def\nilc{{\tt NILC}}
\def\sevem{{\tt SEVEM}}
\def\commander{\texttt{Commander}}
\def\quilc{{\tt QUILC}}
\def\pilc{{\tt PILC}}
\def\prilc{{\tt PRILC}}
\def\ebilc{{\tt EBILC}}

\def\pitf{{\tt PITF}}
\def\pritf{{\tt PRITF}}
\def\quitf{{\tt QUITF}}
\def\ebitf{{\tt EBITF}}
\newcommand{\cmark}{\ding{51}}%
\newcommand{\xmark}{\ding{55}}%
\title[Exploring two-spin linear combinations]{Exploring two-spin internal linear combinations for the recovery of the CMB polarization}

\author[R. Fern\'andez-Cobos et al.]{R. Fern\'andez-Cobos$^1$\thanks{e-mail:cobos@ifca.unican.es}, A. Marcos-Caballero $^1$$^,$$^2$, P. Vielva$^1$, E. Mart\'inez-Gonz\'alez$^1$, \newauthor R. B. Barreiro$^1$\\
$^1$     Instituto de F\'isica de Cantabria, CSIC-Universidad de Cantabria, Avda. de los Castros s/n, E-39005 Santander, Spain. \\
$^2$     Dpto. de F\'isica Moderna, Universidad de Cantabria, Avda. los Castros s/n, E-39005 Santander, Spain.}

\date{Accepted  Received ; in original form }

\begin{document}

\maketitle

\begin{abstract}
We present a methodology to recover cosmic microwave background (CMB) polarization in which the quantity $P = Q+ iU$ is linearly combined at different frequencies using complex coefficients. This is the most general linear combination of the $Q$ and $U$ Stokes parameters which preserves the physical coherence of the residual contribution on the CMB estimation. The approach is applied to the internal linear combination (ILC) and the internal template fitting (ITF) methodologies. The variance of $P$ of the resulting map is minimized to compute the coefficients of the linear combination. One of the key aspects of this procedure is that it serves to account for a global frequency-dependent shift of the polarization phase. Although in the standard case, in which no global E-B transference depending on frequency is expected in the foreground components, minimizing $\left\langle |P|^2\right\rangle$ is similar to minimizing $\left\langle Q^2\right\rangle$ and $\left\langle U^2\right\rangle$ separately (as previous methodologies proceed), multiplying $Q$ and $U$ by different coefficients induces arbitrary changes in the polarization angle and it does not preserve the coherence between the spinorial components. The approach is tested on simulations, obtaining a similar residual level with respect to the one obtained with other implementations of the ILC, and perceiving the polarization rotation of a toy model with the frequency dependence of the Faraday rotation.
\end{abstract}
\section{Introduction}
\label{sec:Introduction}
As it is still a loose thread within the standard model \citep{BICEP2Keck2015, BICEP2Planck2015}, the quest for primordial gravitational waves from inflation stands as a major aim for forthcoming cosmic microwave background (CMB) polarization experiments. If exist, it is widely known that their imprint should be visible in the CMB as B-mode polarization at large scales \citep{Polnarev1985}. The expected CMB polarization is faint with respect to the polarized emission from Galactic foregrounds \citep[see e.g.,][]{Tucci2005}. In addition, the data accuracy which will be provided by incoming experiments makes the CMB recovering more sensitive to the foreground characterization \citep[e.g.,][]{Remazeilles2015}. Therefore, the confluence of these reasons makes the component separation and the recovery of the CMB very important intermediate steps towards the detection of the primordial B-mode anisotropies. 

There is a wide range of component separation methods described in the literature. On the one hand, some component separation methodologies are able to recover several contributions at once. These approaches allow one to obtain all the components assumed to be present in the data, as long as a physical model is provided for each one. Examples of these methods are those based on independent component analysis, such as FastICA or \smica\ \citep[see e.g.,][]{Maino2002, Stivoli2006, Cardoso2008}; maximum entropy \citep[MEM; see e.g.][]{Barreiro2004, Stolyarov2002}; generalized internal linear combinations \citep{Remazeilles2011}; or parametric estimations, such as \commander\ \citep{Eriksen2008}. 

On the other hand, there is a whole set of methods focused on recovering only a particular component from the data whose frequency dependence is known, which is, indeed, the only physical assumption taken into account here about the sky emissions. Although in cosmological analyses this component is typically the CMB, the methodologies can be easily adapted to extract other contributions with a given frequency dependence such as the thermal Sunyaev-Zel'dovich effect, or the galactic molecular CO emission \citep[see e.g.,][]{Hurier2013, Remazeilles2013}. In this category, we find the standard internal linear combination \citep[ILC; e.g.][]{Tegmark1998, Eriksen2004}, in which the foreground removal is performed as a weighted average of the different channels at a common resolution. The internal template fitting (denoted now on by ITF) approach is a particular case of ILC in which some channels are used to build templates for the foreground contamination. 

Within this latter category, both the ILC and the ITF approaches are used to remove the foreground contribution from CMB maps in current experiments. For instance, the \textit{Planck} Collaboration \citep[see][]{PlanckIX2015} uses \nilc\ \citep{Delabrouille2009}, an ILC which works in a given wavelet (needlet) space minimizing the variance of the $E$ and $B$ maps at each scale, and an ITF in real space (\sevem) working directly on the $Q$ and $U$ maps \citep[this latter approach was also implemented in a wavelet space in][]{FernandezCobos2012}.

Due to the anisotropic nature of the foreground contributions, assuming weights for combining globally the whole sky-coverage of each frequency map is not the most efficient way to combine the information. The coefficient estimation would be dominated by the most contaminated regions. On the one hand, to avoid this inconvenience, the ILC approach can be used within different regions of the sky as applied, for instance, by the \textit{WMAP} Collaboration \citep[see e.g.,][]{Gold2011}. However, splicing afterwards different regions on the foreground-reduced map is not a trivial issue. \citet{Park2007} also proposed an ILC application within hundreds of pixel groups with similar foreground espectral indices. On the other hand, the implementation of these methodologies in a wavelet space also allows an effective spatial variation of the coefficients \citep[e.g.][]{FernandezCobos2012}. In addition, another way to consider a scalar variation is to apply the ILC in harmonic space \citep[e.g.][]{Tegmark2003}. A combined approach which takes into account the spatial variation and performs the minimization in harmonic space was proposed by \citet{Kim2008}.

As the $Q$ and $U$ Stokes parameter maps are spinorial components, working on the $E$ and $B$ maps is the most straightforward way to extend the temperature methodology to polarization data, because $E$ and $B$ are scalars as $T$ (in fact, strictly speaking, $B$ is a pseudoscalar). However, as only a partial sky-coverage of data is commonly available from a realistic experiment, deriving the E- and B-mode maps is not a trivial task. In this paper, we present a generalization of the ILC methodology applied directly on the $Q$ and $U$ maps treated as spinorial components. This approach is based on covariant quantities and then preserves the coherence of the spinorial description. As the foreground residuals in forthcoming experiments could be at the same level, or higher, as the CMB signal to be measured, it would be crucial to model the residual component present on the resulting map in order to detect and characterize properly the primordial B-mode polarization. 

This paper is structured as follows: the spinorial methodology is presented in Section~\ref{sec:methodology}, including a review of the standard temperature application. We discuss about the properties of the new proposal in Section~\ref{sec:dis}. An assessment of the methodology with multifrequency simulations is presented in Section~\ref{sec:sim}. Finally, the conclusions are summarized in Section~\ref{sec:Conclusions}.
    
%
 
\section{Methodology}
\label{sec:methodology}
In this section, we present a methodology based on linear combinations which allows one to deal properly with spinorial components. First of all, we review the standard ILC and ITF approaches used in CMB temperature data. Secondly, applications of our spinorial frame to these methodologies are developed. 

As mentioned above, the ILC is the simplest way to perform an internal linear combination in  real space, which consists in a weighted linear combination of different frequency maps. This method is focused on recovering a specific component. In the case of the CMB studies, the primordial fluctuations are the most interesting signal, which is expected to be constant in thermodynamic temperature units in the microwave frequency range. This is the only assumption about the physical properties of the different sky emissions which is needed here. 

The ITF approach is a particular case of ILC with implicit constraints imposed by the construction of the templates as a subtraction of different frequencies at the same resolution. In particular, the coefficient associated with the channel to be cleaned is fixed to $1$. For a template which is built as the substraction of two different channels in order to remove the CMB component, we are assuming that the coefficients associated with each one are equal with opposite sign.

When the foreground removal is performed in real space, the ILC methodology requires that all channels are considered at a common resolution. In contrast, the ITF approach preserves the original resolution of the map to be cleaned, but the foreground removal is conditioned by the effective resolution of the templates. For both methodologies, the real space implementation also allows one to deal with any partial sky-coverage without introducing any systematic effect from the mask.

\subsection{Standard temperature implementation} 
\label{subsec:methodt}
Let us review the standard implementations of the ILC and the ITF approaches for CMB temperature data. In real space, both approaches are based on a linear combination to build the CMB estimation from a multifrequency set of maps. 

On the one hand, within the ILC approach, the CMB signal is estimated as:
\begin{equation}
\hat{T}_{\mathrm{CMB}} = \sum_{i=1}^{N_{\nu}}{\omega_i T_i},
\label{eq:ILCcomb}
\end{equation}
where $T_i$ denotes the corresponding map of frequency $\nu_i$, and $\omega_i$ is a set of $N_{\nu}$ (number of frequency bands) coefficients which are estimated by minimizing the variance of the resulting map.

To guarantee that the CMB component is unbiasedly recovered, we must assume that:
 \begin{equation}
 \label{eq:constraintT}
\sum_{i=1}^{N_{\nu}}{\omega_i} = 1.
\end{equation}

The variance of the resulting map can be written as
\begin{equation}
\left\langle \hat{T}_{\mathrm{CMB}}^2(p) \right\rangle - \left\langle \hat{T}_{\mathrm{CMB}}(p) \right\rangle^2 = \boldsymbol{\omega}^T\mathbf{C}\boldsymbol{\omega}.
\end{equation}
where the angle brackets denote the average over all pixels $p$ in the map, $\boldsymbol{\omega}$ is a column vector with all the coefficients $\omega_i$, and the covariance matrix $\mathbf{C}$ can be expressed as
\begin{equation}
C_{ij} = \left\langle T_i(p) T_j(p) \right\rangle - \left\langle T_i(p) \right\rangle \left\langle T_j(p) \right\rangle.
\end{equation}

Using the method of Lagrange multipliers, the set of coefficients is found solving the linear system of $N_{\nu}$ derivative equations and the constraint given in Equation~(\ref{eq:constraintT}):
\begin{equation}
\left(\begin{array}{cc} 2\mathbf{C} & -\mathbf{1} \\
\mathbf{1}^T & 0 \end{array}\right) \left(\begin{array}{c} \boldsymbol{\omega} \\ \lambda \end{array} \right) = \left(\begin{array}{c} \mathbf{0} \\ 1 \end{array} \right),
\end{equation}
where $\lambda$ is the Lagrange multiplier, and $\mathbf{1}$ and $\mathbf{0}$ denote column arrays of ones and zeros, respectively. The coefficients of the linear combination are finally determined as:
\begin{equation}
\omega_i = \dfrac{\displaystyle \sum_{j=1}^{N_{\nu}}{C^{-1}_{ij}}}{\displaystyle \sum_{i,j=1}^{N_{\nu}}{C^{-1}_{ij}}}.
\label{eq:coeffT_ILC}
\end{equation}

On the other hand, the CMB estimation within the ITF approach is computed as a subtraction between the map which will be cleaned ($d$) and a linear combination of a set of $N_t$ templates ($t_i$):
\begin{equation}
\hat{T}_{\mathrm{CMB}} = d - \sum_{i=1}^{N_t}{\alpha_i t_i},
\end{equation}
where $\alpha_i$ denotes the weight for the template $t_i$. The new configuration space has less degrees of freedom than in the ILC approach, by definition, since $N_t < N_{\nu}$. Note that these coefficients are conceptually different from those shown above for the case of the ILC, but they can be trivially derived from each other by imposing the corresponding constraints. These weights are computed, as in the ILC, by minimizing the variance of the resulting map:
\begin{align}
\left\langle \hat{T}_{\mathrm{CMB}}^2(p) \right\rangle - \left\langle \hat{T}_{\mathrm{CMB}}(p) \right\rangle^2 = & \left\langle d^2(p)\right\rangle - \left\langle d(p)\right\rangle^2 -2 \mathbf{b}\boldsymbol{\alpha} \nonumber \\
 & + \boldsymbol{\alpha}^T\boldsymbol{\Sigma}\boldsymbol{\alpha},
\end{align}
where the covariance matrix $\boldsymbol{\Sigma}$ of the templates can be expressed as:
\begin{equation}
\Sigma_{ij} = \left\langle t_i(p) t_j(p) \right\rangle - \left\langle t_i(p) \right\rangle \left\langle t_j(p) \right\rangle,
\end{equation}
and the vector $\mathbf{b}$ contains information about the correlation between the data and the templates: $b_i \equiv \left\langle t_i(p) d(p)\right\rangle$.

Finally, it is trivial to show that the $N_t$ coefficients can be computed as:
\begin{equation}
\boldsymbol{\alpha} = \boldsymbol{\Sigma}^{-1}\mathbf{b}.
\end{equation}

In the ITF approach, no additional constraints must be considered because they are implicit in the generation of the templates as subtraction of different frequency maps at the same resolution.

\subsection{Polarization ILC}
\label{subsec:ILCpol}
Respecting the philosophy of the ILC, a natural extension of this methodology to the case of CMB polarization, without resort to the E and B modes, is to combine the quantity $Q \pm iU$, which transforms like two-spin variables with $s=\pm2$ under rotations around the local axis defined by the corresponding direction in the sky. Hereafter, we denote this approach as Polarization ILC (\pilc). In this manner, the CMB estimation can be written as:
\begin{equation}
\hat{Q}_{\mathrm{CMB}}(p)\pm i\hat{U}_{\mathrm{CMB}}(p) = \sum_{j=1}^{N_{\nu}}{\left[ \omega^{(R)}_j\pm i\omega^{(I)}_j\right] \left[ Q_j(p) \pm iU_j(p)\right]},
\end{equation}
where $Q_j$ and $U_j$ are the corresponding Stokes parameter maps at frequency $\nu_j$. In this case, the coefficients of the linear combination are complex numbers in such a way that $\omega^{(R)}_j$ and $\omega^{(I)}_j$ denote the real and imaginary parts of these numbers, respectively.

To ensure that the CMB component is unbiasedly recovered, the following constraints must be considered:
\begin{eqnarray}
\sum_{j=1}^{N_{\nu}}{\omega^{(R)}_j} & = & 1,\label{eq:consR} \\
\sum_{j=1}^{N_{\nu}}{\omega^{(I)}_j} & = & 0. \label{eq:consI} 
\end{eqnarray}
A particular scenario in which the Eq.~(\ref{eq:consI}) is satisfied is that where $\omega^{(I)}_j$ is null for all frequencies. Hereafter, we refer to this particular case with only non-zero real coefficients as Polarization Real ILC \citep[\prilc; a similar approach was already described in][]{Kim2009}. 

In the case of the ILC application to CMB temperature described in Section~\ref{subsec:methodt}, 
the estimation of the coefficients is made by minimizing the variance of the resulting map. However, in the polarization case, we deal with components of the polarization vector projected in local frames, and, therefore, $Q$ and $U$ cannot be considered globally. Whilst the mean value of the temperature field can be estimated as an average over all the pixels, it is not possible to proceed in this way in the case of individual spinorial components, because $\left\langle Q \right\rangle$ and $\left\langle U \right\rangle$ cannot be properly estimated. In addition, subtracting a constant contribution from $Q$ or $U$ is equivalent to introduce a pattern on the E and B modes, which depends on the particular coordinate frame used to describe the polarization. In particular, for parity reasons, subtracting a constant from $Q$ induces an E-mode contribution in the even multipoles and a B-mode contribution in the odd multipoles. Conversely, subtracting a constant from the $U$ map induces an E-mode contribution in the odd multipoles and a B-mode contribution in the even multipoles. Therefore, adding a constant to $Q$ and $U$ is not a covariant transformation.

For all these considerations, then, for the polarization case, we choose to minimize a covariant quantity, $\left\langle |\hat{P}_{\mathrm{CMB}}|^2 \right\rangle$, where $\hat{P}_{\mathrm{CMB}} = \hat{Q}_{\mathrm{CMB}}+i\hat{U}_{\mathrm{CMB}}$:
{\small
\begin{align}
 & \left\langle \left[\hat{Q}_{\mathrm{CMB}}(p) + i \hat{U}_{\mathrm{CMB}}(p)\right]\left[\hat{Q}_{\mathrm{CMB}}(p) - i \hat{U}_{\mathrm{CMB}}(p)\right] \right\rangle  \nonumber \\ 
&= \left( \begin{array}{cc} \left[\boldsymbol{\omega}^{(R)}\right]^T & \left[\boldsymbol{\omega}^{(I)}\right]^T\end{array} \right) \left( \begin{array}{cc} \mathbf{C}^{(+)} & -\mathbf{C}^{(-)} \\ 
\mathbf{C}^{(-)} & \mathbf{C}^{(+)} \end{array}\right) \left( \begin{array}{c}\boldsymbol{\omega}^{(R)} \\
\boldsymbol{\omega}^{(I)}\end{array} \right),
\end{align}}
where the components of the matrices $\mathbf{C}^{(+)}$ and $\mathbf{C}^{(-)}$ are:
\begin{eqnarray}
C_{kl}^{(+)} & \equiv & \left\langle Q_k(p) Q_l(p) + U_k(p) U_l(p) \right\rangle, \\
C_{kl}^{(-)} & \equiv & \left\langle Q_k(p) U_l(p) - U_k(p) Q_l(p) \right\rangle. 
\end{eqnarray}
In this context, this quantity becomes the optimal option because
the usual estimator of the variance of $\hat{P}_{\mathrm{CMB}}$: $S(\hat{P}_{\mathrm{CMB}}) = \left\langle \hat{P}_{\mathrm{CMB}} \hat{P}_{\mathrm{CMB}}^*\right\rangle - \left\langle \hat{P}_{\mathrm{CMB}}\right\rangle \left\langle \hat{P}_{\mathrm{CMB}}^*\right\rangle$ is not defined, since the estimation of the expected value of $\hat{P}_{\mathrm{CMB}}$ from the data, as a pixel average, is not covariant.

The linear system of equations compound by the $2N_{\nu}$ minimization conditions and the two constraints can be written as:
\begin{equation}
\left( \begin{array}{cccc} 2\mathbf{C}^{(+)} & -2\mathbf{C}^{(-)} & -\mathbf{1} & \mathbf{0} \\
2\mathbf{C}^{(-)} & 2\mathbf{C}^{(+)} & \mathbf{0} & -\mathbf{1} \\
\mathbf{1}^T & \mathbf{0}^T & 0 & 0 \\
\mathbf{0}^T & \mathbf{1}^T & 0 & 0 \end{array} \right) \left(\begin{array}{c} \boldsymbol{\omega}^{(R)} \\
\boldsymbol{\omega}^{(I)} \\
\lambda_R  \\
\lambda_I \end{array}\right) = \left(\begin{array}{c} \mathbf{0} \\
\mathbf{0} \\
1  \\
0 \end{array}\right),
\end{equation}
where $\lambda_R$ and $\lambda_I$ denote the Lagrange multipliers.

Therefore, the coefficients can be computed as:
\begin{eqnarray}
\omega^{(R)}_k & = & \dfrac{\lambda_R}{2} \sum_{l=1}^{N_{\nu}}{C^{-1}_{kl}} + \dfrac{\lambda_I}{2} \sum_{l=N_{\nu}+1}^{2N_{\nu}}{C^{-1}_{kl}},  \\
\omega^{(I)}_k & = & \dfrac{\lambda_R}{2} \sum_{l=1}^{N_{\nu}}{C^{-1}_{N_{\nu}+k,l}} + \dfrac{\lambda_I}{2} \sum_{l=N_{\nu}+1}^{2N_{\nu}}{C^{-1}_{N_{\nu}+k,l}},
\end{eqnarray}
where 
\begin{equation}
\mathbf{C} \equiv \left( \begin{array}{cc}
\mathbf{C}^{(+)} & -\mathbf{C}^{(-)} \\
\mathbf{C}^{(-)} & \mathbf{C}^{(+)}
\end{array} \right).
\end{equation}
Finally, solving for the Lagrange multipliers, and imposing the constraints from Eq.~(\ref{eq:consR}) and (\ref{eq:consI}), it is obtained:
\begin{eqnarray}
\dfrac{\lambda_R}{2} & = & \dfrac{S_{+}}{S_{+}^2-S_{-}^2}, \\
\dfrac{\lambda_I}{2} & = & \dfrac{-S_{-}}{S_{+}^2-S_{-}^2},
\end{eqnarray}
where 
\begin{eqnarray}
S_{+} & \equiv & \sum_{i,j=1}^{N_{\nu}}{C^{-1}_{ij}}, \\
S_{-} & \equiv & \sum_{i=1}^{N_{\nu}}{\sum_{j=N_{\nu}+1}^{2N_{\nu}}{C^{-1}_{ij}}}.
\end{eqnarray}

The terms of the covariance matrix $\mathbf{C}$ guarantee that all quantities involved in the minimization are independent with respect to the polarization local frame.

Note that, if the terms of $\mathbf{C}^{(-)}$ are negligible, we obtain the particular case of \prilc, in which the $\omega^{(I)}_k$ coefficients vanish and the expression for the coefficients in the temperature case presented in Equation~(\ref{eq:coeffT_ILC}), with $\mathbf{C} = \mathbf{C^{(+)}}$ (equivalent to minimize jointly $\left\langle \hat{Q}^2_{\mathrm{CMB}} + \hat{U}^2_{\mathrm{CMB}} \right\rangle$; i.e., $\left\langle |\hat{P}_{\mathrm{CMB}}|^2\right\rangle$ with only a set of $N_{\nu}$ coefficients), is recovered for the real part $\omega^{(R)}_k$:
\begin{equation}
\omega^{(R)}_k = \dfrac{\displaystyle \sum_{l=1}^{N_{\nu}}{\left(C^{(+)}\right)^{-1}_{kl}}}{\displaystyle \sum_{k,l=1}^{N_{\nu}}{\left(C^{(+)}\right)^{-1}_{kl}}}.
\label{eq:coeffpol_wi}
\end{equation}

\subsection{Polarization ITF}
We generalize the ITF approach for the case of CMB polarization in the same way followed for the ILC. In this case (from now on called Polarization ITF; \pitf), the CMB estimator can be written as:
\begin{align}
\hat{Q}_{\mathrm{CMB}}(p)\pm i\hat{U}_{\mathrm{CMB}}(p) = & \left[d^{(Q)}(p) \pm i d^{(U)}(p)\right] \nonumber \\
 &- \sum_{j=1}^{N_{\nu}}{\left[ \alpha^{(R)}_j\pm i\alpha^{(I)}_j\right] \left[ t^{(Q)}_j(p) \pm it^{(U)}_j(p)\right]},
\end{align}
where $d^{(Q)}(p)$ and $d^{(U)}(p)$ denote the polarization components of the data map to be cleaned, whilst $t^{(Q)}_j(p)$ and $t^{(U)}_j(p)$ represent the Stokes parameters of the template $t_j$, at the pixel $p$. 

As in the ILC case, the expected value of $|\hat{P}_{\mathrm{CMB}}|^2$ is the quantity chosen to be minimized to obtain the complex coefficients $\boldsymbol{\alpha}$:
\begin{align} 
\left\langle \hat{P}^2_{\mathrm{CMB}}(p) \right\rangle = & \left\langle \left[d^{(Q)}\right]^2\right\rangle + \left\langle \left[d^{(U)}\right]^2\right\rangle \nonumber \\
& + \left( \begin{array}{cc} \left[\boldsymbol{\alpha}^{(R)}\right]^T & \left[\boldsymbol{\alpha}^{(I)}\right]^T\end{array} \right) \left[ \left(\begin{array}{c} -2\mathbf{b}^{(+)} \\
-2\mathbf{b}^{(-)} \end{array}\right)\right. \nonumber \\ 
& + \left. \left( \begin{array}{cc} \boldsymbol{\Sigma}^{(+)} & -\boldsymbol{\Sigma}^{(-)} \\ 
\boldsymbol{\Sigma}^{(-)} & \boldsymbol{\Sigma}^{(+)} \end{array}\right) \left( \begin{array}{c}\boldsymbol{\alpha}^{(R)} \\
\boldsymbol{\alpha}^{(I)}\end{array} \right) \right],
\end{align}
where 
\begin{eqnarray}
\Sigma_{ij}^{(+)} & \equiv & \left\langle t^{(Q)}_i(p) t^{(Q)}_j(p) + t^{(U)}_i(p) t^{(U)}_j(p) \right\rangle \\
\Sigma_{ij}^{(-)} & \equiv & \left\langle t^{(Q)}_i(p) t^{(U)}_j(p) - t^{(U)}_i(p) t^{(Q)}_j(p) \right\rangle, 
\end{eqnarray}
and 
\begin{eqnarray}
b^{(+)}_i & \equiv &  \left\langle t^{(Q)}_i(p)d^{(Q)}(p) + t^{(U)}_i(p)d^{(U)}(p) \right\rangle \\
b^{(-)}_i & \equiv & \left\langle t^{(Q)}_i(p)d^{(U)}(p) - t^{(U)}_i(p)d^{(Q)}(p) \right\rangle.
\end{eqnarray}
When the minimization condition is imposed, we obtain the following linear system of $2N_t$ equations:
\begin{equation}
\left( \begin{array}{cc} \boldsymbol{\Sigma}^{(+)} & -\boldsymbol{\Sigma}^{(-)} \\
\boldsymbol{\Sigma}^{(-)} & \boldsymbol{\Sigma}^{(+)} \end{array} \right) \left(\begin{array}{c} \boldsymbol{\alpha}^{(R)} \\
\boldsymbol{\alpha}^{(I)} \end{array}\right) = \left(\begin{array}{c} \mathbf{b}^{(+)} \\
\mathbf{b}^{(-)} \end{array}\right).
\end{equation}
Therefore, the final expressions for the coefficients are:
\begin{eqnarray}
\alpha^{(R)}_i & = & \sum_{j=1}^{N_t}{\Sigma^{-1}_{ij} b^{(+)}_j} + \sum_{j=N_t+1}^{2N_t}{\Sigma^{-1}_{ij} b^{(-)}_j}, \\
\alpha^{(I)}_i & = & \sum_{j=1}^{N_t}{\Sigma^{-1}_{i+N_t,j} b^{(+)}_j} + \sum_{j=N_t+1}^{2N_t}{\Sigma^{-1}_{i+N_t,j} b^{(-)}_j},
\end{eqnarray}
where 
\begin{equation}
\boldsymbol{\Sigma} \equiv \left( \begin{array}{cc}
\boldsymbol{\Sigma}^{(+)} & -\boldsymbol{\Sigma}^{(-)} \\
\boldsymbol{\Sigma}^{(-)} & \boldsymbol{\Sigma}^{(+)}
\end{array} \right).
\end{equation}

As in Section~\ref{subsec:ILCpol}, when the $\boldsymbol{\Sigma}^{(-)}$ contribution is assumed to vanish, and $\mathbf{b}^{(-)}$ is neglected, we recover the same expression for the $\boldsymbol{\alpha}^{(R)}$ coefficients than the one obtained for the temperature case, except for the fact that the covariance matrix and the correlations between templates and data have to be jointly taken for $Q$ and $U$, as minimizing $\left\langle \hat{Q}^2_{\mathrm{CMB}} + \hat{U}^2_{\mathrm{CMB}} \right\rangle$:
\begin{equation}
\boldsymbol{\alpha}^{(R)} = \left(\boldsymbol{\Sigma}^{(+)}\right)^{-1}\mathbf{b}^{(+)}.
\end{equation}
For completeness, we denote this particular case as Polarization Real ITF (\pritf).

\section{Properties of the two-spin cleaned map}
\label{sec:dis}
In this section, we stand out the peculiarities of our proposal for CMB polarization. As the ITF can be seen as a particular case of the ILC, for simplicity, let us focus the discussion on the latter approach. We pay attention to two important aspects related to our proposal: the possibility of preserving the physical interpretation of the residuals, and the role played by the $\omega^{(I)}_j$ coefficients.

As it was discussed above, the \pilc\ approach is the most general way to extend the standard ILC to polarization data combining the multifrequency set of quantities $(Q_j\pm i U_j)$. An alternative approach could be working on each one of the Stokes parameters independently. Here, $Q$ and $U$ at each frequency $\nu_j$ are weighted, as in the case of temperature data (see Section~\ref{subsec:methodt}), by different real coefficients $\omega^{(Q)}_j$ and $\omega^{(U)}_j$, respectively. Hereafter, this approach is denoted as \quilc, and it is followed, for instance, within the \sevem\ methodology by the \textit{Planck} Collaboration, although based on the ITF. Its standing point is based on considering the Stokes parameter maps as independent scalar images, such that the CMB signal is preserved. However, we remind that this procedure is not covariant. Although the quasi-variances of $\hat{Q}_{\mathrm{CMB}}$ and $\hat{U}_{\mathrm{CMB}}$ are independently minimized in \citet{PlanckIX2015}, the results from \quilc\ in this paper are computed by minimizing the expected values of $\hat{Q}^2_{\mathrm{CMB}}$ and $\hat{U}^2_{\mathrm{CMB}}$ for a direct comparison with \pilc\ and \prilc. Nevertheless, the results from both estimators are very similar.

\subsection{Keeping the physical interpretation of the residuals}
\label{subsec:phys_int}
In the case of microwave polarization data, it is important to keep the coherence between the spinorial components $Q$ and $U$. Actually, this should be satisfied not only for the CMB (which is guaranteed in \pilc, \prilc\ and \quilc, by construction), but also for the foreground residual, which, in many cases (e.g., for building an estimation from the angular power spectrum), has to be physically modelled. We show below that both \pilc\ and \prilc\ imply a proper treatment of the spinor, and they allow one to deal physically with the residual contribution.

In terms of the spin-weighted spherical harmonics ${}_{s}Y_{\ell m}$, the quantity $Q\pm iU$ at the direction $\mathbf{n}$ can be expanded as \citep[see, e.g.][]{Zaldarriaga1997}:
\begin{equation}
\left(Q\pm iU\right)(\mathbf{n}) = \sum_{\ell=2}^{\infty}{\sum_{m=-\ell}^{\ell}{a^{\pm2}_{\ell m}\ {}_{\pm2}Y_{\ell m}(\mathbf{n})}},
\end{equation}
where the coefficients of the expansion $a^{\pm2}_{\ell m}$ are related to the E- and B-mode polarization spherical harmonics, $e_{\ell m}$ and $b_{\ell m}$, as: $a^{\pm 2}_{\ell m} = e_{\ell m} \pm ib_{\ell m}$.

The polarization spherical harmonic coefficients of the cleaned CMB can therefore be written as:
\begin{equation}
\label{eq:almP}
\left(\begin{array}{c}
\hat{e}^{(\mathrm{CMB})}_{\ell m} \\
\hat{b}^{(\mathrm{CMB})}_{\ell m} 
\end{array} \right) = \sum_{j=1}^{N_{\nu}}{\left( \begin{array}{cc}
\omega^{(R)}_j & -\omega^{(I)}_j \\
\omega^{(I)}_j & \omega^{(R)}_j
\end{array} \right) \left( \begin{array}{c}
e^{j}_{\ell m} \\
b^{j}_{\ell m}
\end{array} \right)}, 
\end{equation}
where $e^{j}_{\ell m}$ and $b^{j}_{\ell m}$ are the E- and B-mode polarization spherical harmonics of the corresponding channel at $\nu_j$. 

When the \prilc\ approach is considered, all the $\boldsymbol{\omega}^{(I)}$ coefficients vanish and the combination is performed with the real coefficients $\boldsymbol{\omega}^{(R)}$ (see Eq.~\ref{eq:coeffpol_wi}). As mentioned, this is equivalent to minimize $\left\langle Q^2_{\mathrm{CMB}} + U^2_{\mathrm{CMB}}\right\rangle$ (i.e., $\left\langle|\hat{P}_{\mathrm{CMB}}|^2 \right\rangle$) with a unique weight for $Q$ and $U$ at each frequency.
 
On the other hand, within the \quilc\ approach, the quantities which are imposed to be minimal, $\left\langle \hat{Q}^2_{\mathrm{CMB}}\right\rangle$ and $\left\langle \hat{U}^2_{\mathrm{CMB}} \right\rangle$ respectively, depends on the local polarization frame. In this sense, \quilc\ spoils the physical meaning of the residual component. The residual contribution of the resulting map has not a proper physical description and it should be considered as a mere residual of the signal processing. In addition, weighting $Q$ and $U$ at each frequency $\nu_j$ by different real coefficients $\omega^{(Q)}_j$ and $\omega^{(U)}_j$ introduces a new term with respect to the standard situation. To show that, we expand the quantity $\left[\omega^{(Q)}_jQ_j + i \omega^{(U)}_jU_j\right](\mathbf{n})$ in terms of the spin-weighted spherical harmonics ${}_{\pm2}Y_{\ell m}$:
\begin{eqnarray}
\label{eq:harqu}
\left[\omega^{(Q)}_jQ_j + i \omega^{(U)}_jU_j\right](\mathbf{n}) & = & \sum_{\ell=2}^{\infty}\sum_{m=-\ell}^{\ell}\left[\mu_j \left(a^{+2}_{\ell m}\right)^j\ {}_{+2}Y_{\ell m}(\mathbf{n}) \right. \nonumber \\
& & \left. + \eta_j \left(a^{-2}_{\ell m}\right)^j\ {}_{-2}Y_{\ell m}(\mathbf{n}) \right],  \\
\label{eq:harqu2}
\left[\omega^{(Q)}_jQ_j - i \omega^{(U)}_jU_j\right](\mathbf{n}) & = & \sum_{\ell=2}^{\infty}\sum_{m=-\ell}^{\ell}\left[ \eta_j \left(a^{+2}_{\ell m}\right)^j\ {}_{+2}Y_{\ell m}(\mathbf{n}) \right. \nonumber \\
& & \left. + \mu_j \left(a^{-2}_{\ell m}\right)^j\ {}_{-2}Y_{\ell m}(\mathbf{n}) \right], 
\end{eqnarray}
where $\left(a^{\pm 2}_{\ell m}\right)^j$ are the spin-weighted spherical harmonic coefficients corresponding to the frequency $\nu_j$, and $\mu_j \equiv \left[\omega^{(Q)}_j+\omega^{(U)}_j\right]/2$, and $\eta_j \equiv \left[\omega^{(Q)}_j-\omega^{(U)}_j\right]/2$. In terms of these coefficient combinations, the constraint imposed by the \quilc\ approach, equivalent to that in Eq.~(\ref{eq:consR}) applied separately to $\boldsymbol{\omega}^{(Q)}$ and $\boldsymbol{\omega}^{(U)}$, leads to $\sum \mu_j = 1$ and $\sum \eta_j = 0$. These constraints are the reason why the CMB component is preserved in the \quilc\ methodology. 

In the case in which $\omega^{(Q)}_j= \omega^{(U)}_j$, the contribution which involves $\eta$ vanishes at each frequency, and the E- and B-mode spherical harmonics can be computed from $\omega^{(Q)}_j\left[Q_j \pm i U_j\right]$ as usual. This is an approach similar to \prilc\ but minimizing quantities that are not covariant. However, in the generic case in which $\omega^{(Q)}_j \ne \omega^{(U)}_j$, the $\eta$-term introduces a four-spin contribution in the residuals of the resulting maps when the spin raising and lowering operators $\eth$ and $\eth^{*}$ are applied to obtain the E- and B-mode polarization fields. Therefore, the resulting $E$ and $B$ are not scalars.

Note that, even in the ideal case in which we have a full-sky coverage and no frequency-dependent leakage components are considered, neither \prilc\ nor \quilc\ with $\omega^{(Q)}_j= \omega^{(U)}_j$ for all frequencies are equivalent to remove the foreground contribution in $E$ and $B$, unless $\left\langle \hat{E}^2_{\mathrm{CMB}} + \hat{B}^2_{\mathrm{CMB}} \right\rangle$ are jointly minimized and a unique coefficient is used to weight both modes, i.e., $\omega^{(E)}_j= \omega^{(B)}_j$, where $\omega^{(E)}_j$ and $\omega^{(B)}_j$ denote the coefficients computed by minimizing independently $\left\langle \hat{E}^2_{\mathrm{CMB}} \right\rangle$ and $\left\langle \hat{B}^2_{\mathrm{CMB}} \right\rangle$. Obviously, as these modes are independent for the CMB component, the ILC estimation can be computed as in the temperature case for the $E$ and $B$ maps as:
\begin{equation}
\left(\begin{array}{c}
\hat{e}^{(\mathrm{CMB})}_{\ell m} \\
\hat{b}^{(\mathrm{CMB})}_{\ell m} 
\end{array} \right) = \sum_{j=1}^{N_{\nu}}{\left( \begin{array}{cc}
\omega^{(E)}_j & 0 \\
0 & \omega^{(B)}_j
\end{array} \right) \left( \begin{array}{c}
e^{j}_{\ell m} \\
b^{j}_{\ell m}
\end{array} \right)}. 
\label{eq:coeffebilc}
\end{equation}
For consistency, we denote this approach as \ebilc. As mentioned, a similar approach applied on needlet space (\nilc) was used by the \textit{Planck} Collaboration \citep{PlanckIX2015}. Let us show that this procedure, in general, introduces a non orientation-preserving term in the residual contribution. It can be shown that, in this case, the spin-weighted spherical harmonic coefficients of the resulting map can be expressed as:
\begin{equation}
\left(a^{+2}_{\ell m}\right)^{(\mathrm{CMB})} = \sum_{j=1}^{N_{\nu}}\left[\mu^{(EB)}_j + \eta^{(EB)}_j \mathcal{P}\right] (a^{+2}_{\ell m})^j,
\end{equation}
where $\mu_j^{(EB)} \equiv \left[\omega^{(E)}_j+\omega^{(B)}_j\right]/2$ and $\eta_j^{(EB)} \equiv \left[\omega^{(E)}_j-\omega^{(B)}_j\right]/2$. The $\eta$-term implies a transformation from $a^{+2}_{\ell m}$ to $a^{-2}_{\ell m}$, which corresponds to a parity transformation $\mathcal{P}$ in the tangent plane\footnote{The tangent plane is that spanned by $\left(\mathbf{e}_{\theta},\ \mathbf{e}_{\phi} \right)$, such that a parity transformation $\mathcal{P}$ implies: $\left(\mathbf{e}_{\theta},\ \mathbf{e}_{\phi} \right) \rightarrow \left(\mathbf{e}_{\theta},\ -\mathbf{e}_{\phi} \right)$.}. From these resulting spherical harmonic coefficients, it is trivial that the Stokes parameter maps of the resulting polarization field can be calculated as:
\begin{equation}
\hat{Q}_{\mathrm{CMB}} + i\hat{U}_{\mathrm{CMB}} = \sum_{j=1}^{N_{\nu}} \left[\mu^{(EB)}_j + \eta^{(EB)}_j \mathcal{P}\right]\left( Q_j + i U_j\right).
\end{equation}
Therefore, although the \ebilc\ approach introduces a $\eta$-term in the $Q$ and $U$ residuals of the resulting maps when $\omega^{(E)}_j \ne \omega^{(B)}_j$, the procedure is still covariant. As in the case of \quilc, these transformations are not propagated to the CMB component because of the constraint $\sum \eta_j = 0$.

Although the instrumental noise and the foreground components introduce a similar contribution in the $E$ and $B$ maps, the CMB signal has different contributions to each mode. It contributes differently to the covariance matrix such that the values of the coefficients for the independent modes $\omega^{(E)}_j$ and $\omega^{(B)}_j$ of a specific realization are affected in a different way by the cross-correlation terms with the CMB component. These contributions tend to be equal  only in the limiting case in which the CMB is negligible with respect to the rest of contributions. In contrast, in the case of \quilc, the coefficients $\omega^{(Q)}_j$ and $\omega^{(U)}_j$ tend to be equal because the variances of the CMB Stokes parameters are comparable to each other.

Summarizing, whilst the \pilc\ and \prilc\ methodologies deal with covariant quantities and are coherent with the physical description of the residuals in the resulting map, this is not the case for the \quilc\ approach, since a non-covariant contribution is artificially included. In the case of \ebilc, the procedure is still covariant, but it introduces a non orientation-preserving term in the residual contribution. An overview of all these methodologies, along with the corresponding ones to the ITF approach, is presented in Table~\ref{tab:methods}.

\begin{table*}
\begin{tabular}{cccccccc}
\hline
\hline
Acronym & Name & DoF & Coefficients & Minimization & Covariant & Rotations & Orientation-preserving \\
\hline
\pilc\ & Polarization ILC & $2N_{\nu}$ & $\boldsymbol{\omega}^{(R)},\ \boldsymbol{\omega}^{(I)}$ & $\left\langle |\hat{P}_{\mathrm{CMB}}|^2 \right\rangle$ & \cmark & \cmark & \cmark \\
\prilc\ & Polarization Real ILC & $N_{\nu}$ & $\boldsymbol{\omega}^{(R)}$ & $\left\langle |\hat{P}_{\mathrm{CMB}}|^2 \right\rangle$ & \cmark & \xmark & \cmark \\
\quilc\ & Q and U ILC & $2N_{\nu}$ & $\boldsymbol{\omega}^{(Q)},\ \boldsymbol{\omega}^{(U)}$ & $\left\langle \hat{Q}_{\mathrm{CMB}}^2 \right\rangle,\ \left\langle \hat{U}_{\mathrm{CMB}}^2 \right\rangle$ & \xmark & - & - \\
\ebilc\ & E and B ILC & $2N_{\nu}$ & $\boldsymbol{\omega}^{(E)},\ \boldsymbol{\omega}^{(B)}$ & $\left\langle \hat{E}_{\mathrm{CMB}}^2 \right\rangle,\ \left\langle \hat{B}_{\mathrm{CMB}}^2 \right\rangle$ & \cmark & \xmark & \xmark \\
\hline
\pitf\ & Polarization ITF & $2N_{t}$ & $\boldsymbol{\alpha}^{(R)},\ \boldsymbol{\alpha}^{(I)}$ & $\left\langle |\hat{P}_{\mathrm{CMB}}|^2 \right\rangle$ & \cmark & \cmark & \cmark \\
\pritf\ & Polarization Real ITF & $N_{t}$ & $\boldsymbol{\alpha}^{(R)}$ & $\left\langle |\hat{P}_{\mathrm{CMB}}|^2 \right\rangle$ & \cmark & \xmark & \cmark \\
\quitf\ & Q and U ITF & $2N_{t}$ & $\boldsymbol{\alpha}^{(Q)},\ \boldsymbol{\alpha}^{(U)}$ & $\left\langle \hat{Q}_{\mathrm{CMB}}^2 \right\rangle,\ \left\langle \hat{U}_{\mathrm{CMB}}^2 \right\rangle$ & \xmark & - & - \\
\ebitf\ & E and B ITF & $2N_{t}$ & $\boldsymbol{\alpha}^{(E)},\ \boldsymbol{\alpha}^{(B)}$ & $\left\langle \hat{E}_{\mathrm{CMB}}^2 \right\rangle,\ \left\langle \hat{B}_{\mathrm{CMB}}^2 \right\rangle$ & \cmark & \xmark & \xmark \\
\hline
\end{tabular}
\caption{Summary of methodologies. From left to right: acronym and complete name of each approach; degrees of freedom (DoF) considered in the minimization, where $N_{\nu}$ and $N_{t}$ are the number of frequencies and templates, respectively; the quantity chosen to be minimal; and other properties which characterize each method, such as the use of covariant quantities, allowing polarization rotations, or the implication of orientation-preserving transformations. Although the discussion in the text is focus on the ILC approach, the corresponding ITF approaches are also included for completeness.}
\label{tab:methods}
\end{table*}

\subsection{Frequency-dependent phase shift}
\label{subsec:phase}
Within the \pilc\ method, the resulting combination allows mixing between E- and B-mode polarization due to the phase of the complex coefficients. Explicitly, the coefficient matrix in Eq.~(\ref{eq:almP}) can be seen as a global rotation of the polarization headless vector by making the change to polar coordinates: 

\begin{equation}
\left( \begin{array}{cc}
\omega^{(R)}_j & -\omega^{(I)}_j \\
\omega^{(I)}_j & \omega^{(R)}_j
\end{array} \right) = |\omega_j|\left( \begin{array}{cc}
\cos\left(2\phi_j\right) & -\sin\left(2\phi_j\right) \\
\sin\left(2\phi_j\right) & \cos\left(2\phi_j\right)
\end{array} \right).
\end{equation}

The phase of the complex coefficient $\phi_j$ is interpreted as the angle of a two-spin global rotation of the spherical harmonic coefficients of E- and B-mode polarization at each frequency $\nu_j$. The presence of this phase shift depends only on a non-null combination of channels: $\left\langle Q_i(p) U_j(p) - U_i(p) Q_j(p) \right\rangle$, for a pair of frequencies $\nu_i$ and $\nu_j$. This implies that the methodology is only sensitive to frequency-dependent changes in the polarization direction. A hypothetical component whose polarization is rotated the same angle over the entire frequency range is innocuous to the method, in the sense that it does not contribute to $\mathrm{\mathbf{C}^{(-)}}$. Therefore, in the case in which we had a component  whose polarization angle suffered a global frequency-dependent shift, we would obtain $\boldsymbol{\omega}^{(I)}\neq 0$.

Within the standard frame, the changes of the polarization angle of the foreground component with frequency are expected to be local, such that, when we consider the pixel average, the effective contributions, such as the global shift expected from the particular configuration of the Galactic magnetic field, seem to be subdominant with respect to the instrumental noise levels. In addition, due to the variation of the foreground polarization modulus with frequency, there should be an induced shift in the effective polarization angle of the total sky emission, but  also subdominant in average. These effective variations, along with any spurious contribution from the specific noise and foreground realizations to the $\mathrm{\mathbf{C}^{(-)}}$ matrix, are fitted by the $\boldsymbol{\omega}^{(I)}$ coefficients, such that the expected value of $|\hat{P}_{\mathrm{CMB}}|^2$ of the resulting map from \pilc\ is smaller than the value obtained from \prilc. In practice, for a particular realization, these coefficients take small values with respect to the $\boldsymbol{\omega}^{(R)}$ ones but non-zero. In this context, when we consider the ensemble average of realizations, the expected value of these coefficients should be close to zero. In contrast, when a global shift of the polarization angle is considered this contribution becomes significant. 

Therefore, the \pilc\ methodology has the potential to be useful to characterize some physical effects, such as the Faraday rotation or other leakages, which induce a polarization phase with frequency dependence. As the Faraday rotation affects to both the CMB and the foregrounds, it should be necessary to isolate the dependence of the CMB in order to obtain an unbiased recovery. In addition, the Faraday rotation depends on the magnetic field. As this phase is taken into account over the whole map, and the changes in the polarization direction are expected to be local, the global effect is expected to be subdominant but, as mentioned, due to the particular shape of the Galactic magnetic field, non-zero. The impact of these coefficients would be more important when the ILC is performed considering different sky regions with coherent magnetic field or spatial variation of the coefficients is allowed, for instance, by implementing the minimization on a wavelet space. In addition, this methodology could be useful for the CMB recovery in scenarios with a frequency-dependent birefringence effect \citep[see, e.g.][]{Gubitosi2014}.

\section{Assessment with simulations}
\label{sec:sim}
To test the \pilc\ approach, we use multifrequency sets of simulations. The frequency range between $45\ \mathrm{GHz}$ and $795\ \mathrm{GHz}$ is chosen as a specific example of a large sky-coverage polarization experiment. In particular, $15$ frequency bands are taken from the Cosmic Origins Explorer (COrE) proposal \citep{Core2011}. As the ILC methodology in real space requires all frequency maps at the same resolution, all channels are considered at a HEALPix resolution \citep{Gorski2005} of $N_{\mathrm{side}} = 256$ and convolved by an effective beam of $\mathrm{FWHM}=25\arcmin$.

As a starting point, we use a CMB fiducial model which accounts only for the B-mode lensing contribution (i.e., a tensor-to-scalar ratio $r=0$). For each frequency $\nu_j$, the polarized foreground contribution is simulated using the Planck Sky Model \citep[PSM;][]{Delabrouille2013}. These realizations account for a synchrotron component, a contribution from thermal dust, and a component due to point sources \citep[fainter than $50\ \mathrm{mJy}$, since the brightest point sources are supposed to be previously removed from the frequency maps, see e.g.,][]{Caniego2009}. Both the specific beam size for each frequency and the nominal instrumental noise levels are shown in Table~\ref{tab:freq}. The \textit{Planck} common mask for polarization analysis is used \citep{PlanckIX2015}.

\begin{table}
\begin{tabular}{ccc}
\hline
\hline
$\nu$ [GHz] & FWHM [arcmin] & Sensitivity [$\mathrm{\mu K \cdot arcmin}$] \\
\hline
45 & 23.0 & 9.07 \\ 
75 & 14.0 & 4.72 \\  
105 & 10.0 & 4.63 \\
135 & 7.8 & 4.55 \\
165 & 6.4 & 4.61 \\
195 & 5.4 & 4.54 \\
225 & 4.7 & 4.57 \\
255 & 4.1 & 10.5 \\
285 & 3.7 & 17.4 \\  
315 & 3.3 & 46.6 \\
375 & 2.8 & 119.0 \\
435 & 2.4 & 258.0 \\
555 & 1.9 & 626.0 \\
675 & 1.6 & 3640.0 \\
795 & 1.3 & 22200.0 \\
\hline  
\end{tabular}
\caption{Specifications of the set of simulations, taken from \citet{Core2011}: centres of the frequency bands, the full width at half maximum (FWHM) of the Gaussian beam of each channel, and the corresponding sensitivity.}
\label{tab:freq}
\end{table}

These simulations are analysed in Section~\ref{subsec:psm}. In addition, we also consider (Section~\ref{subsec:toymodel}) a toy model which accounts for a global polarization rotation of the foreground component, with the frequency dependence presented by the Faraday rotation.

\subsection{Two-spin performance using PSM foregrounds}
\label{subsec:psm}
The residual level of the foreground-reduced maps obtained from the  \quilc, \pilc\ and \prilc\ approaches are quantified in terms of the variance of the resulting $P$ map and the angular power spectra of the E and B modes. As the methodologies are linear, the foreground and the instrumental noise components of each simulation can be propagated through the linear combination, by fixing the coefficients computed from the complete data set (i.e., including the CMB). As the input foreground contribution is the same for all sets of simulations, the randomness which is taken into account comes from the cosmic variance from the CMB signal and the uncertainty due to the instrumental noise. The residuals on the resulting maps depend on the coefficient estimation.

\begin{figure*}
\centering
\includegraphics[scale=0.23]{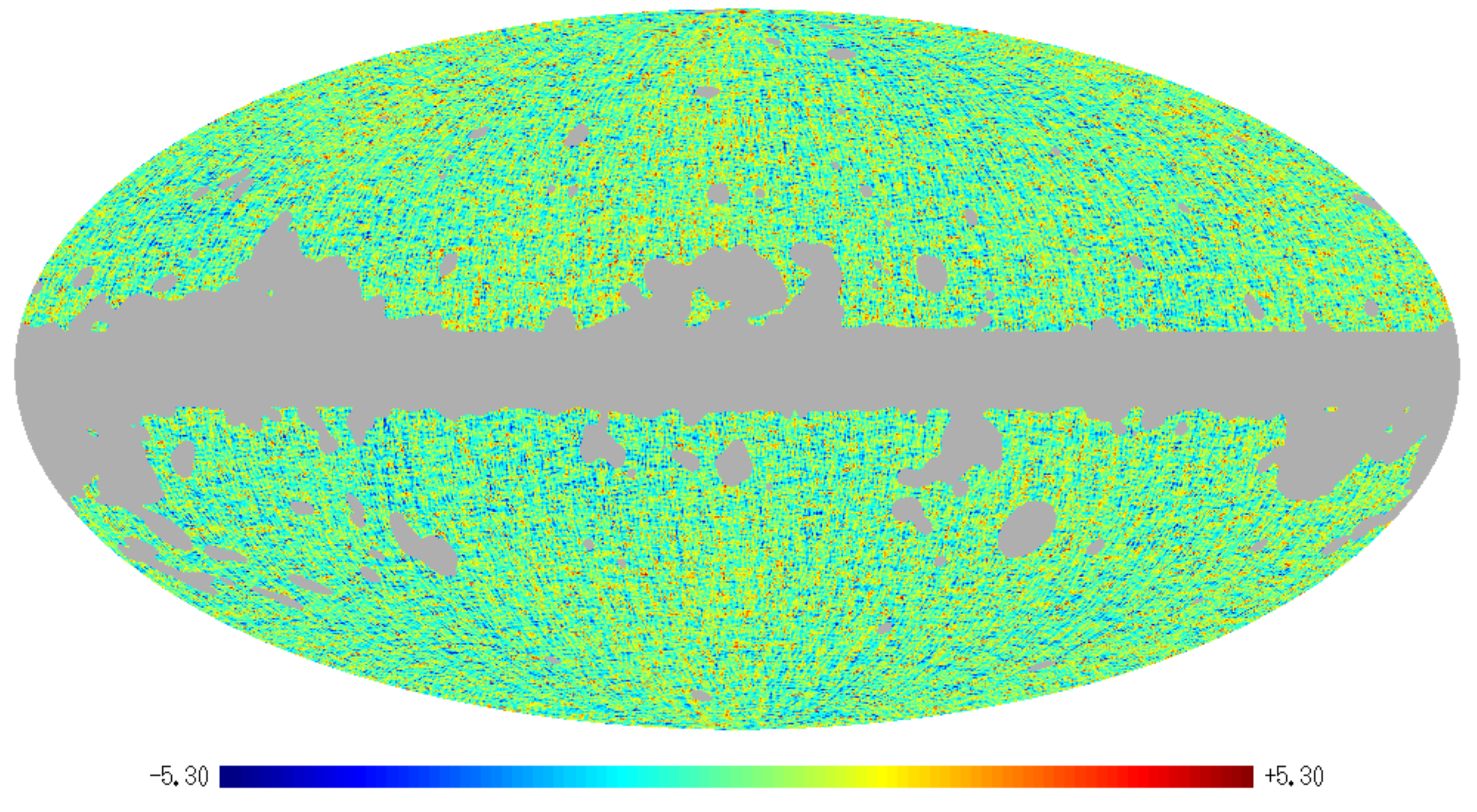}
\includegraphics[scale=0.23]{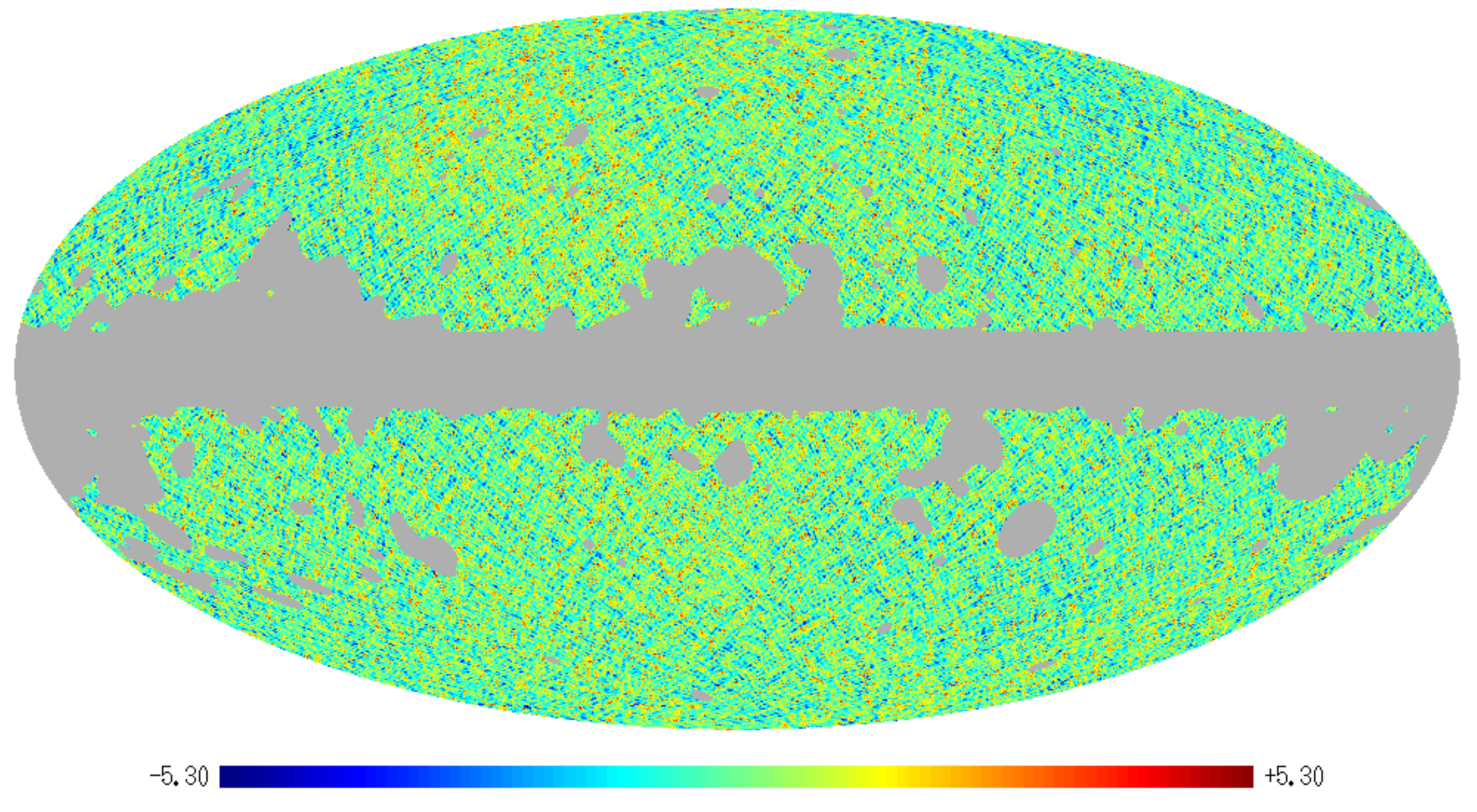} \\
\includegraphics[scale=0.23]{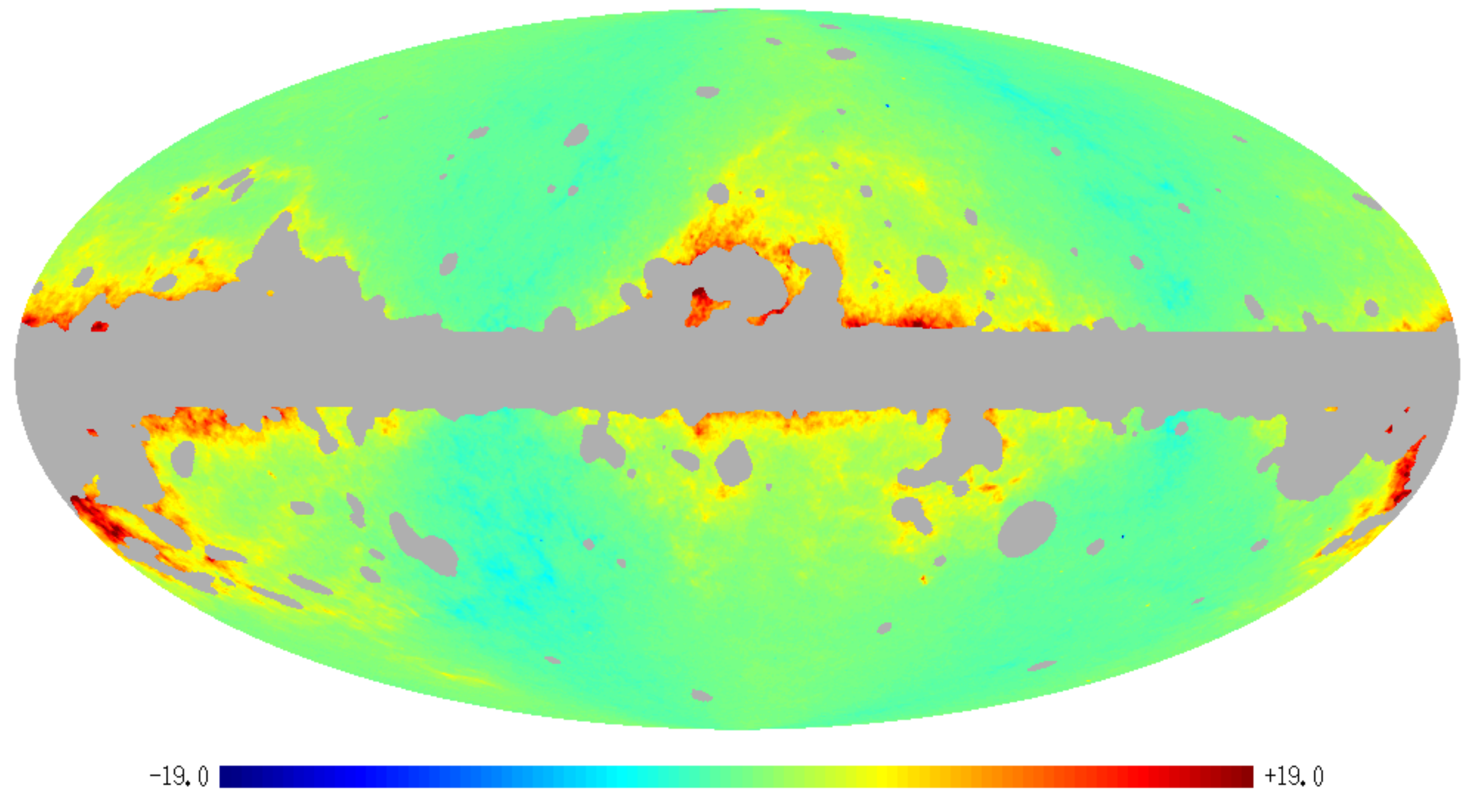}
\includegraphics[scale=0.23]{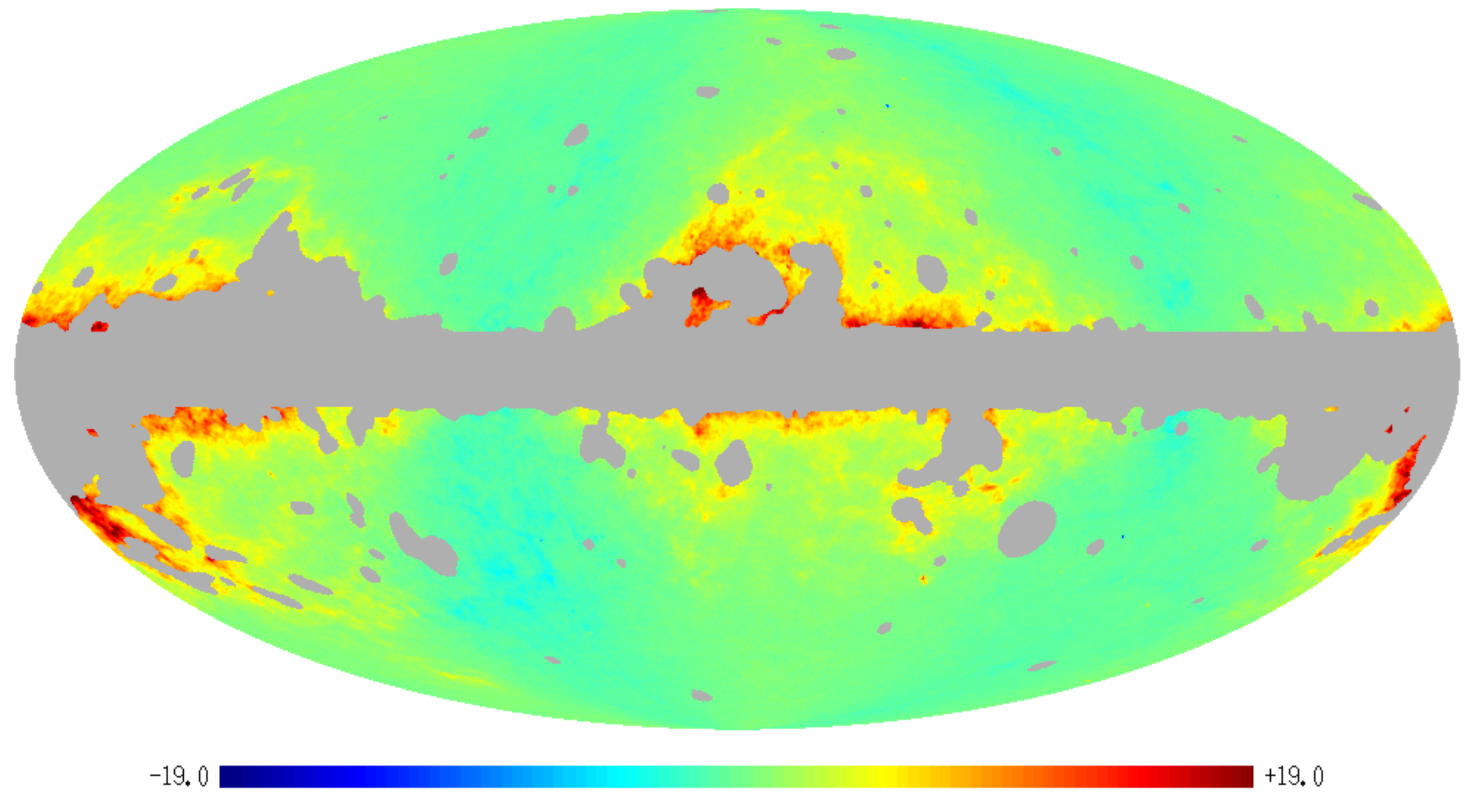} \\
\includegraphics[scale=0.23]{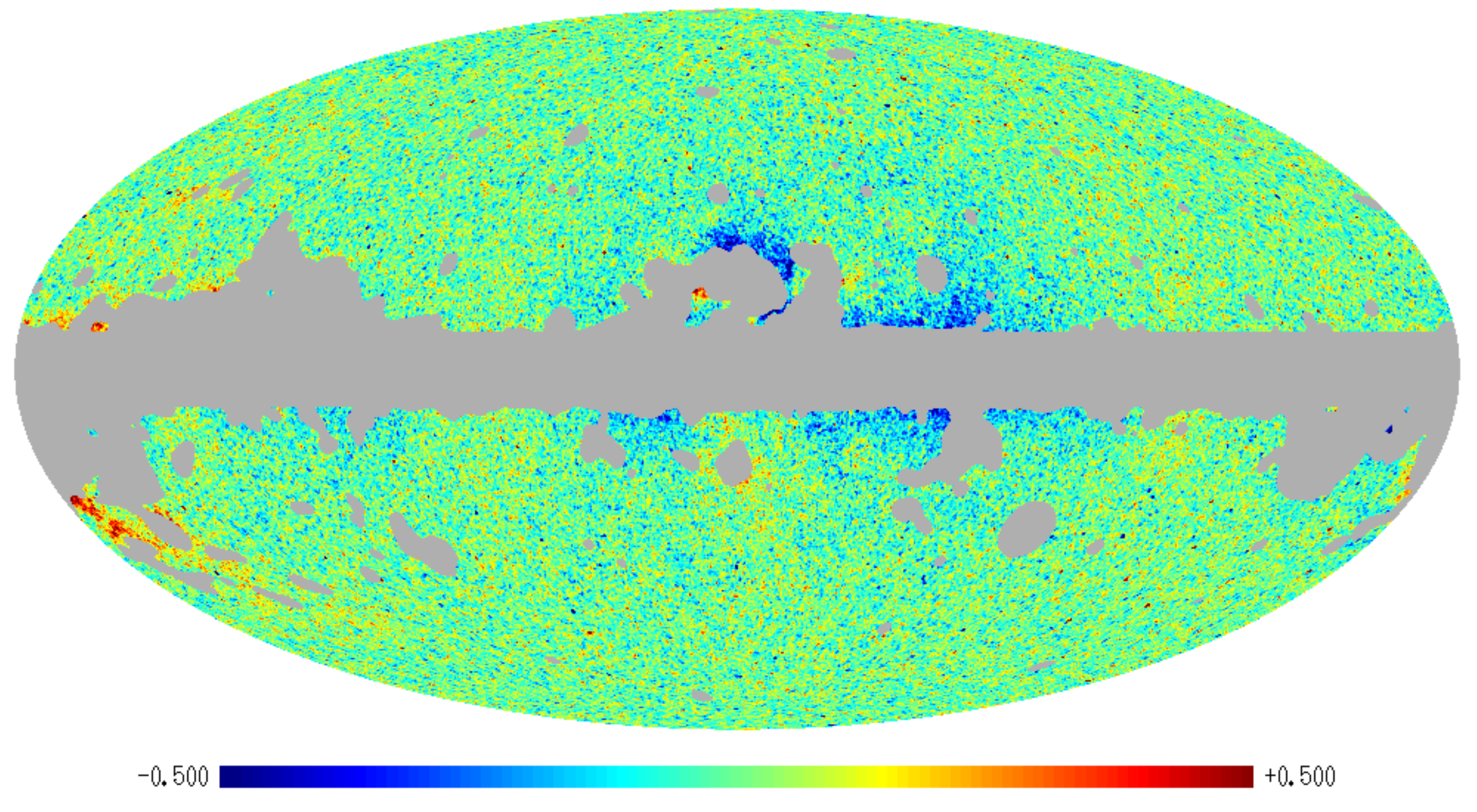}
\includegraphics[scale=0.23]{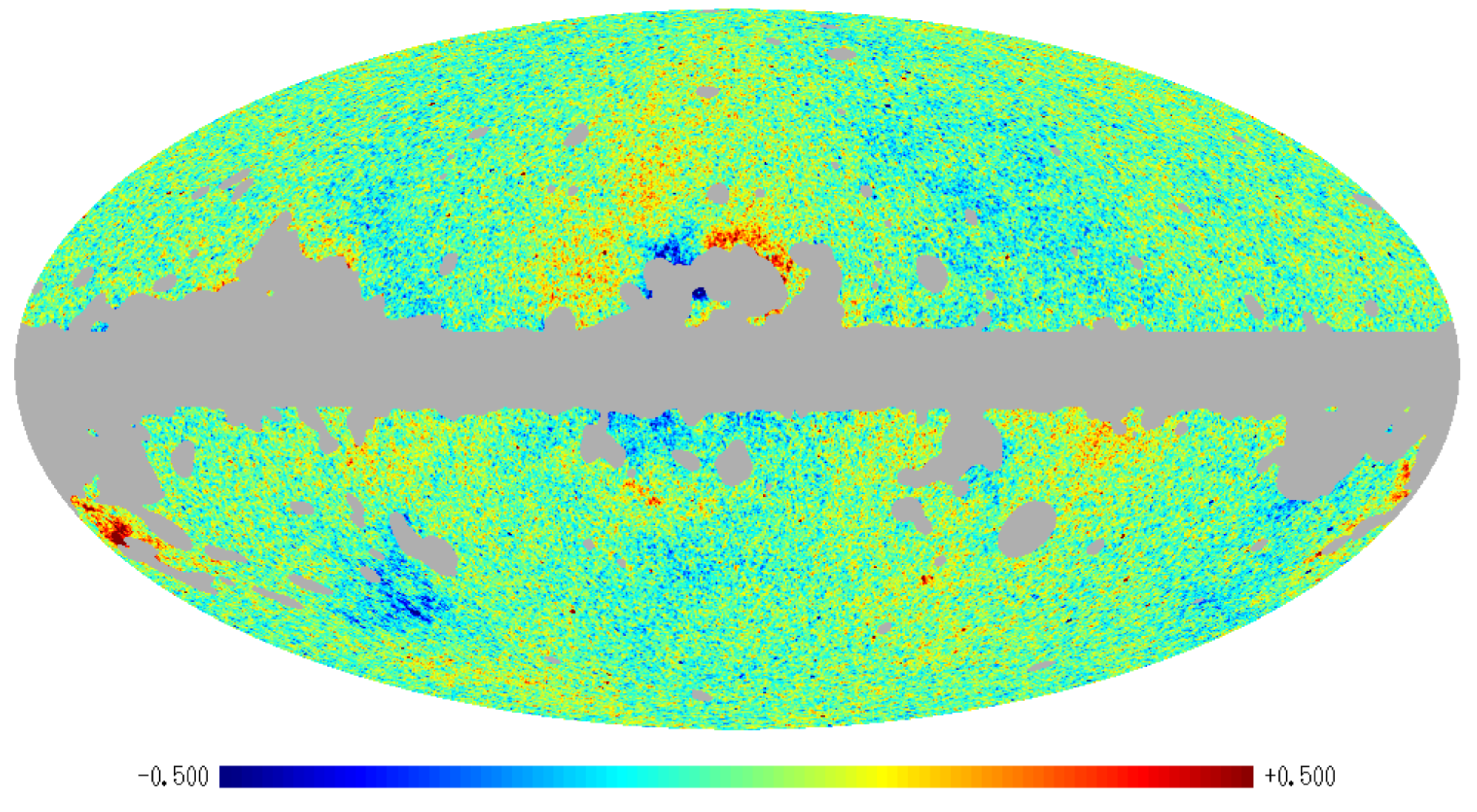} \\
\includegraphics[scale=0.23]{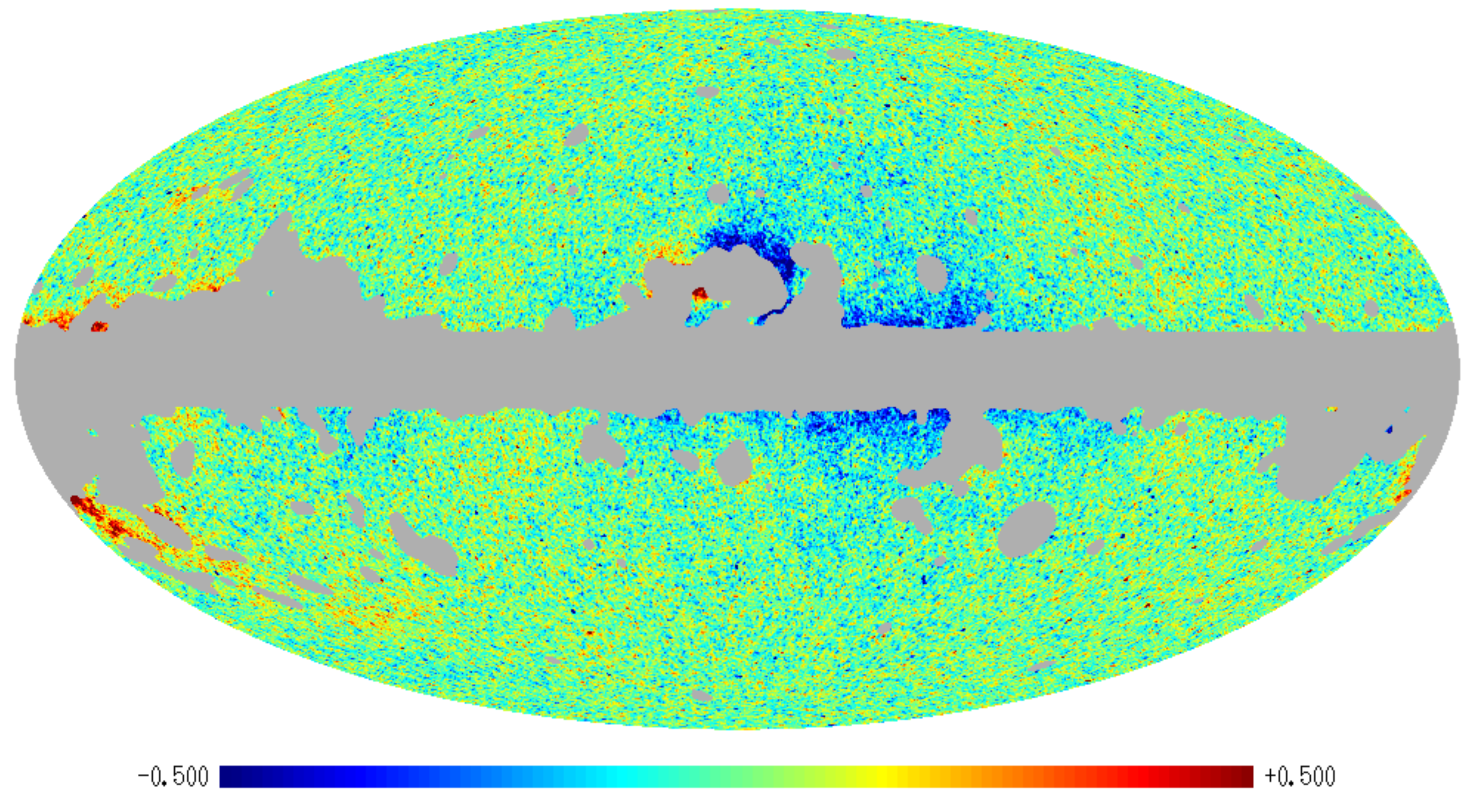}
\includegraphics[scale=0.23]{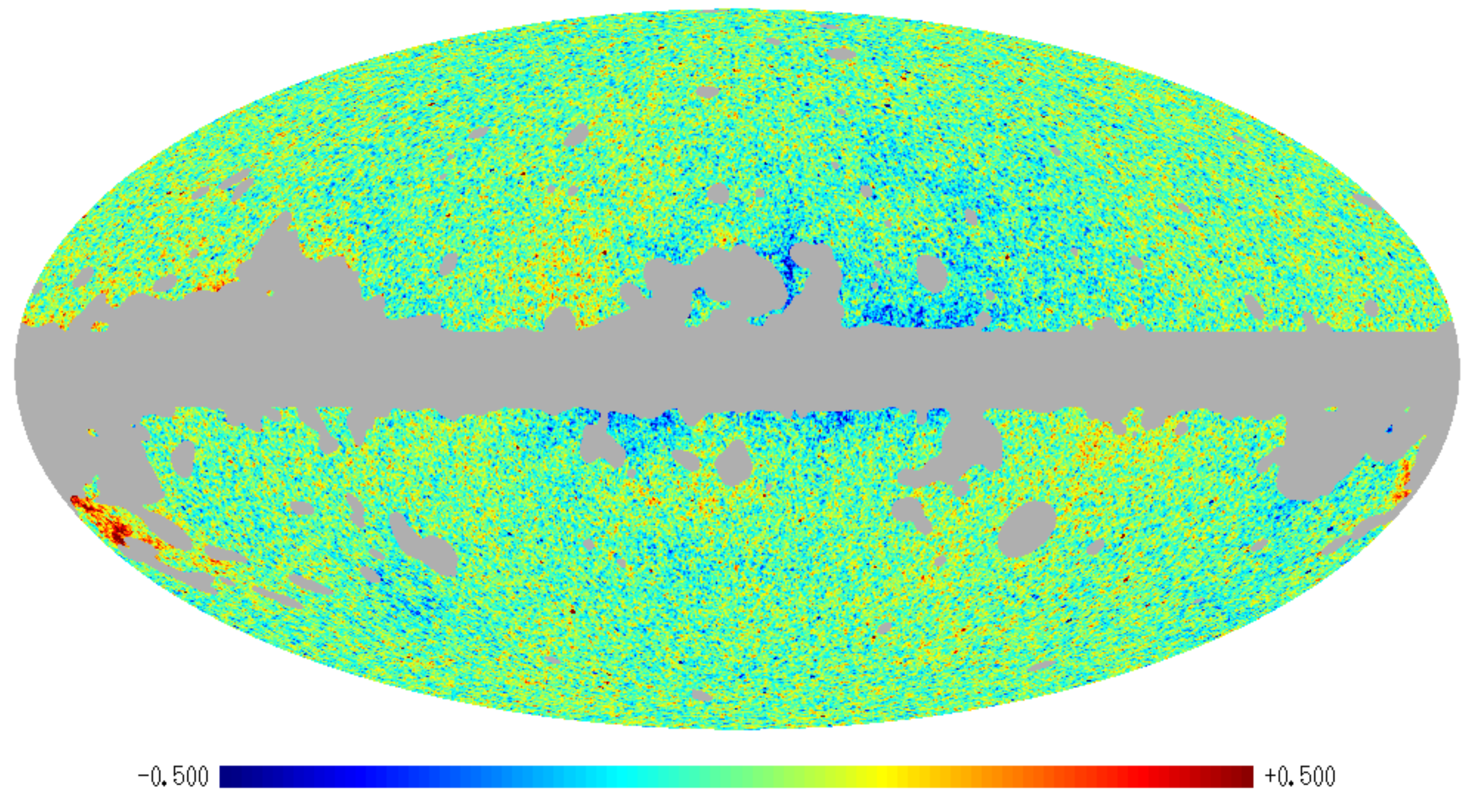} \\
\includegraphics[scale=0.23]{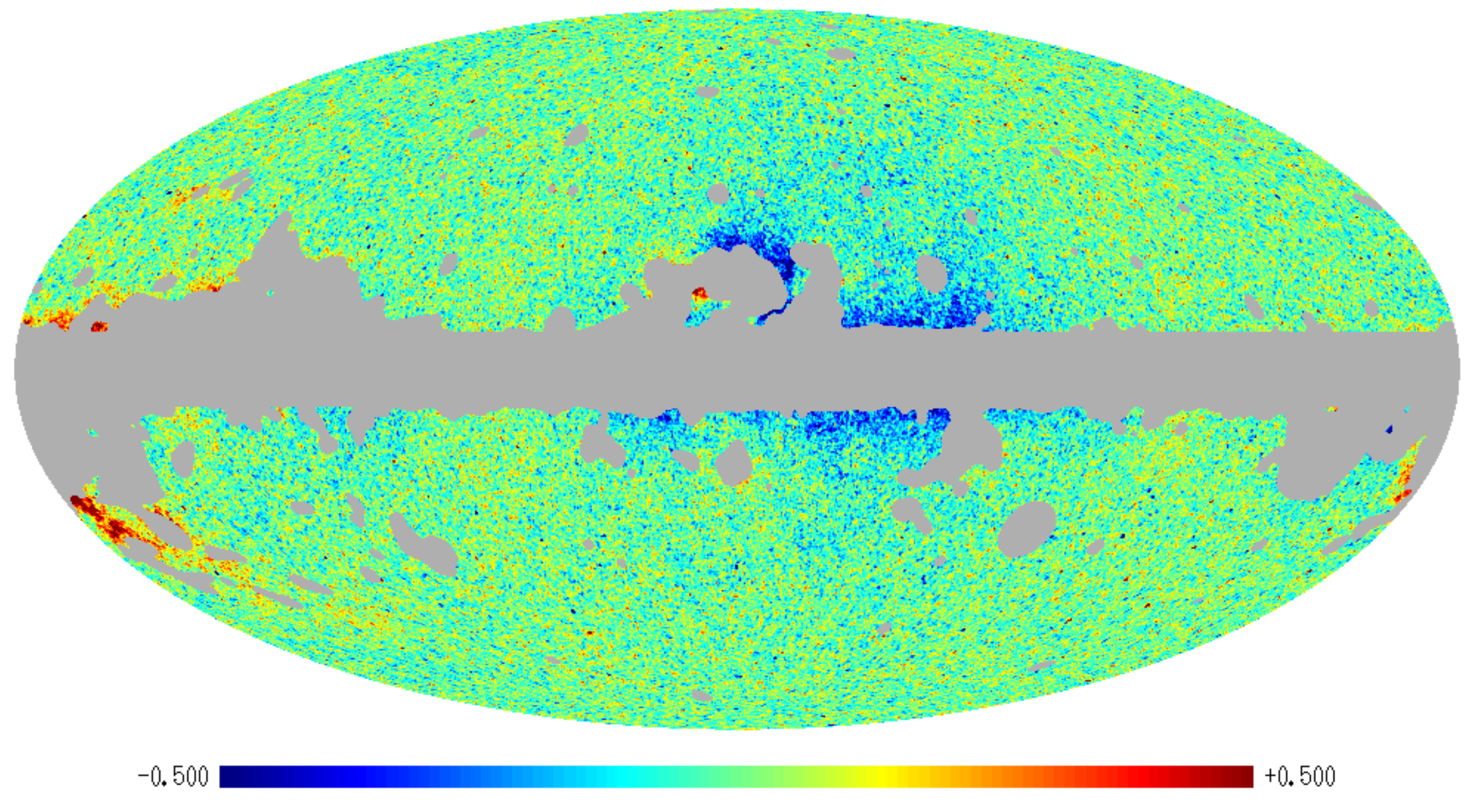}
\includegraphics[scale=0.23]{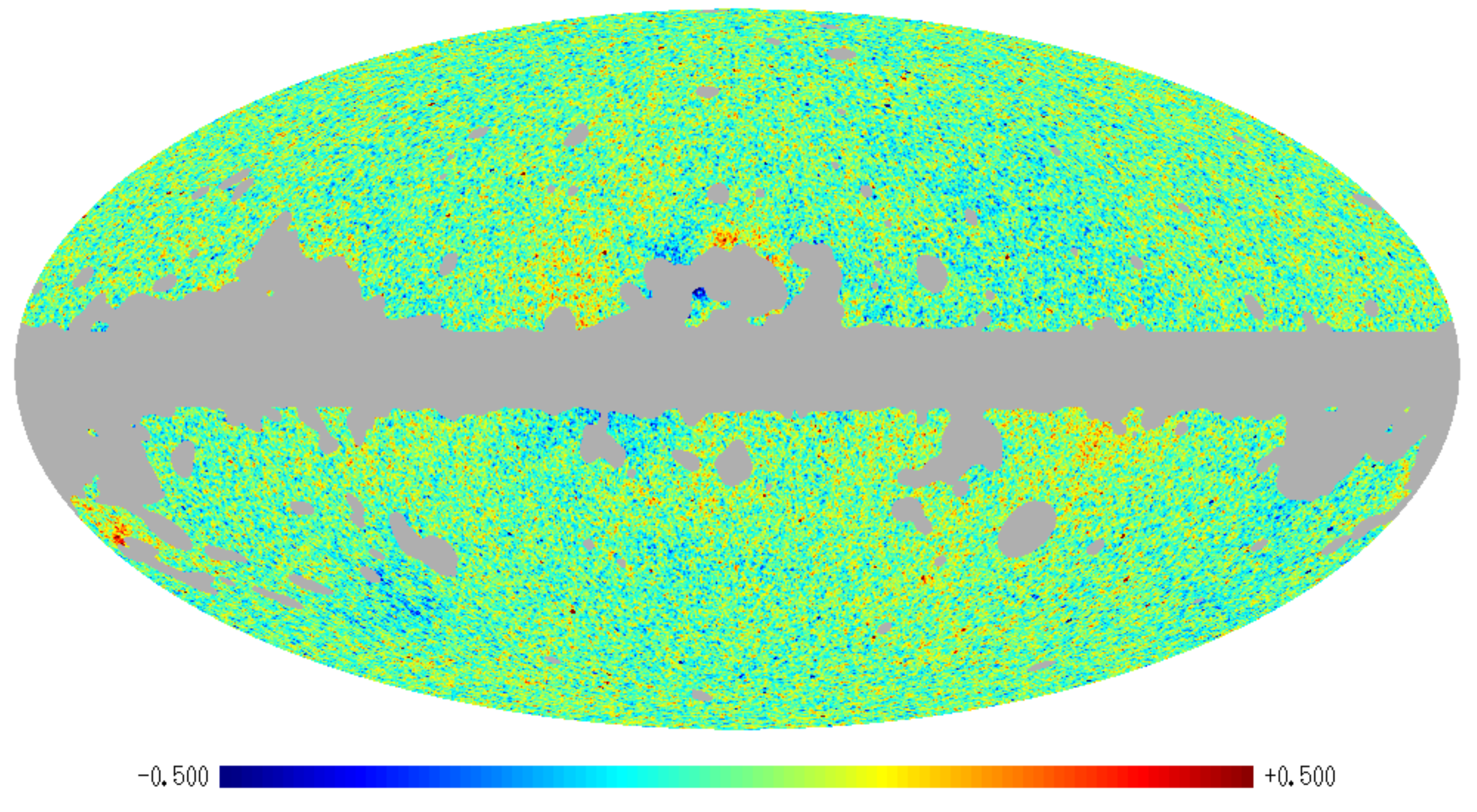} 
\caption{$Q$ (left column) and $U$ (right column) total residual maps from the different ILC polarization approaches for a particular simulation. For comparison, the input CMB realization is shown in the first row, and the input foreground plus noise contribution for the $135\ \mathrm{GHz}$ channel is shown in the second row. The third row corresponds to the total residual maps from \quilc, whilst the fourth and the fifth ones represents the total residuals from \pilc\ and \prilc, respectively.}
\label{fig:maps_res}
\end{figure*}

As the foreground-reduced maps of each methodology are very similar, the corresponding total residual components (from instrumental noise and foregrounds) are shown in Figure~\ref{fig:maps_res} from a random simulation. Some differences between methodologies can be seen close to the Galactic plane. For illustration, the input CMB realization and the total foreground and noise contribution for the $135\ \mathrm{GHz}$ channel are also given in the same figure.

\subsubsection*{Variance of the resulting maps}
\begin{figure}
\centering
\includegraphics[scale=0.35]{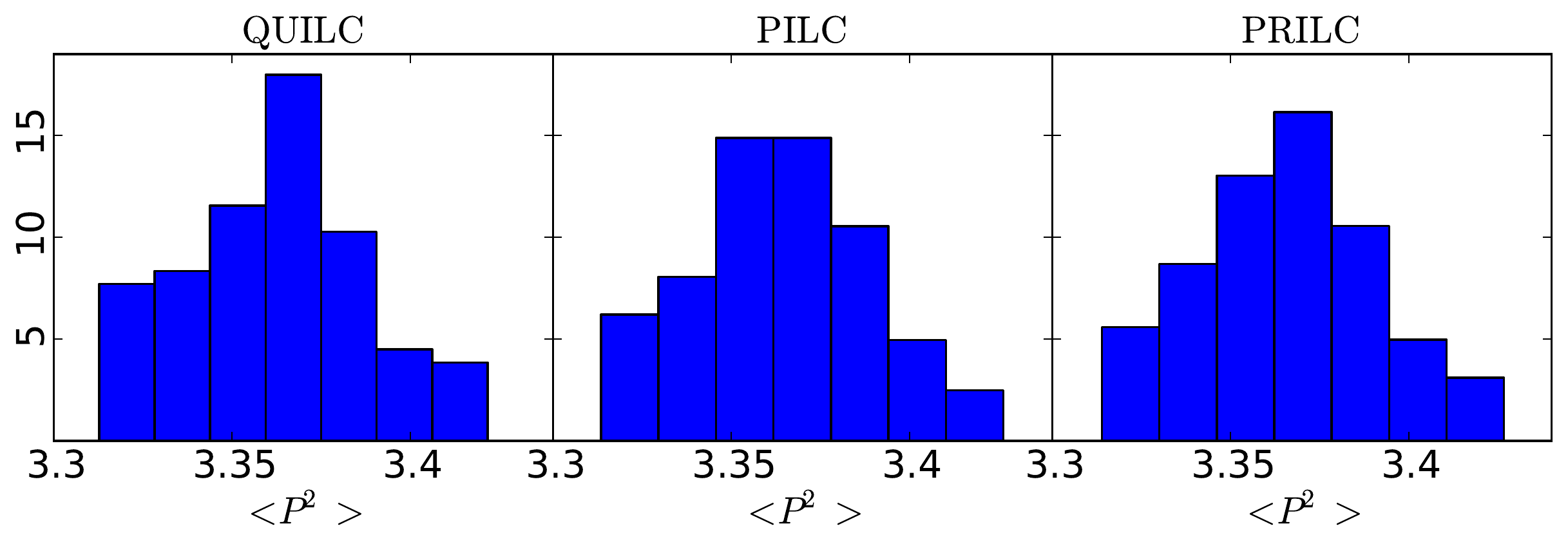}
\includegraphics[scale=0.35]{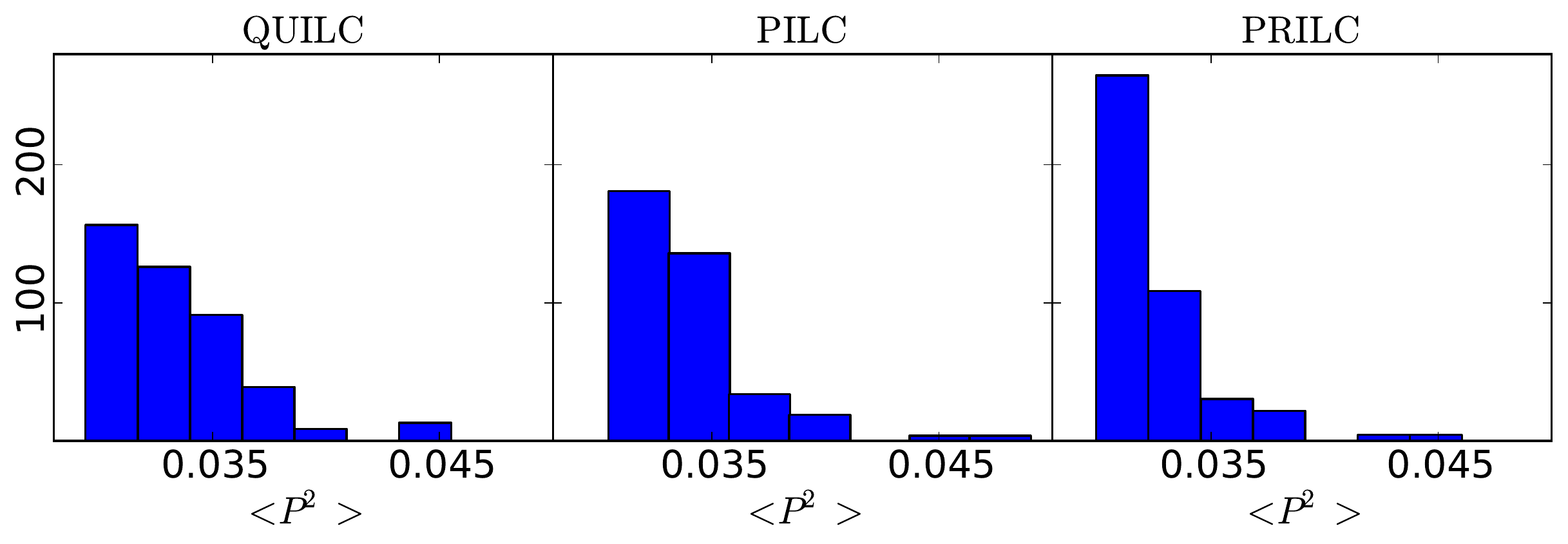}
\includegraphics[scale=0.35]{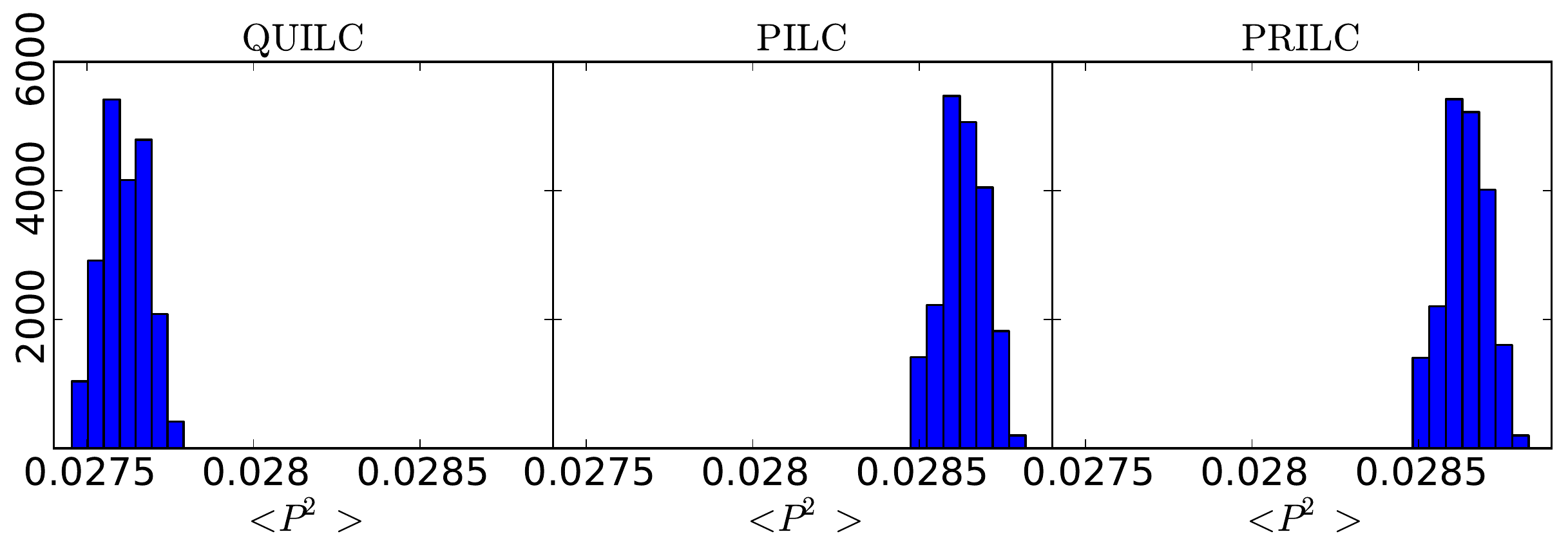}
\caption{Distributions of the mean value of $|\hat{P}_{\mathrm{CMB}}|^2$ estimated from  $100$ sets of simulations with a foreground contribution simulated with the PSM. From left to right, the distributions from \quilc, \pilc\ and \prilc\ are depicted. The top row corresponds to the standard case in which the coefficients are computed from the total map. The contribution from the instrumental noise and foreground residuals in the same case is shown in the middle row.
The ideal case in which the covariance matrices of the foreground and the instrumental noise components are known (instead of using an estimation of the total covariance, as in the standard case, including CMB) is shown in the bottom row.}
\label{fig:variances}
\end{figure}

\begin{figure}
\centering
\includegraphics[scale=0.35]{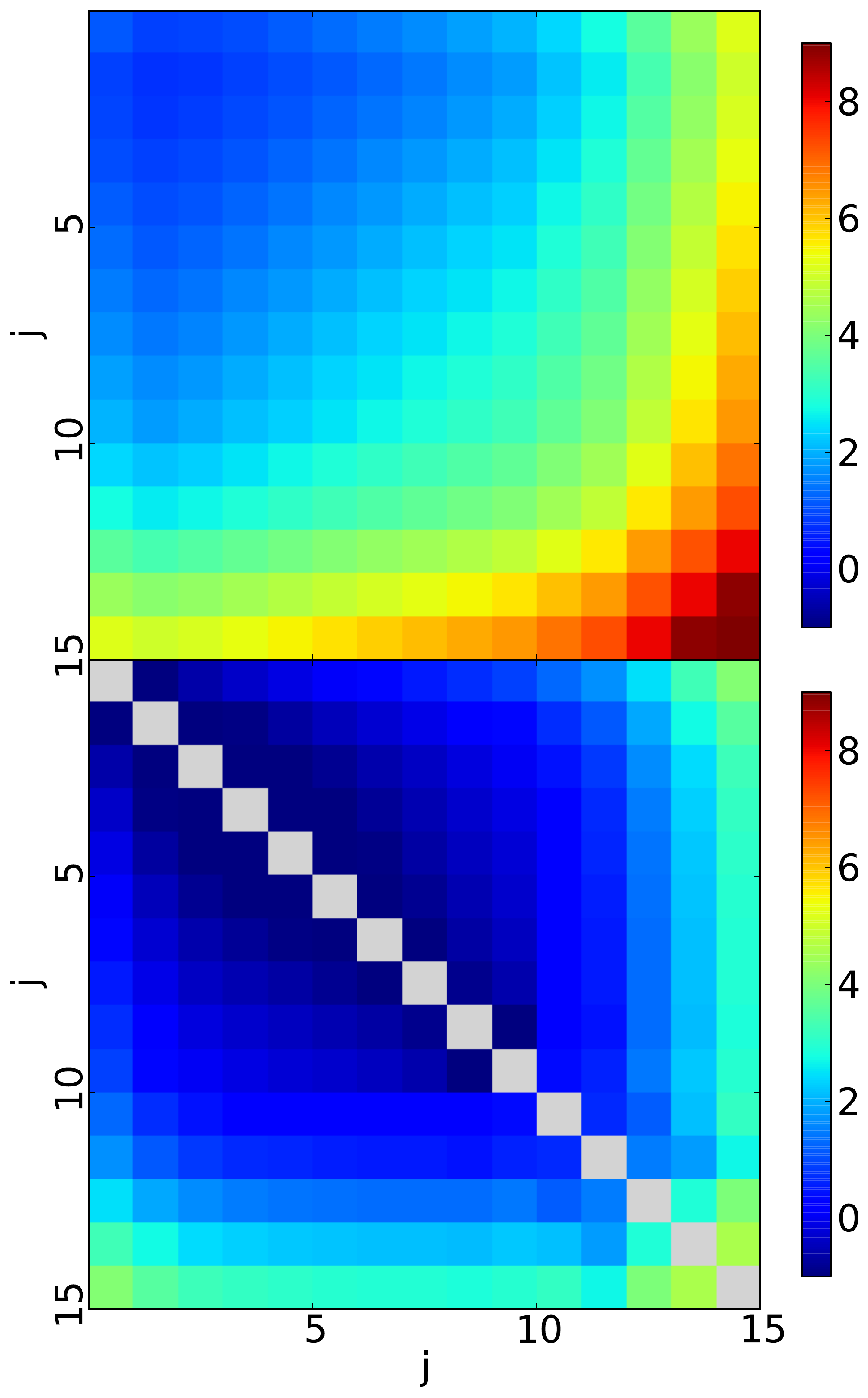}
\caption{Logarithm (base $10$) of the terms of $\mathbf{C}^{(+)}$ (top panel) and the absolute value of $\mathbf{C}^{(-)}$ (bottom panel; in this case the diagonal terms are masked, since they vanish). The axes coordinates correspond to the subscript $j$, denoting the corresponding frequency $\nu_j$.}
\label{fig:matrix}
\end{figure}

In Figure~\ref{fig:variances}, we show the distributions of the mean value of $|\hat{P}_{\mathrm{CMB}}|^2$ estimated from simulations. In the upper row, we depict this distribution for (from left to right) \quilc, \pilc\ and \prilc. The intrinsic variations of $\left\langle |\hat{P}_{\mathrm{CMB}}|^2 \right\rangle$ are very large in comparison with the differences between methodologies due to the cosmic variance. One possibility to measure these differences could be to analyse the variance contribution from only the instrumental noise and foreground residuals (middle row). However, as each simulation provides a set of coefficients conditioned by the cross-correlation between these contributions and the specific CMB realization, this variance is biased \citep[see e.g.,][]{Delabrouille2009}. Therefore, we consider an ideal case (bottom row) in which we could estimate the covariance matrices of the foreground and instrumental noise components instead of using an estimation of the total covariance, including the CMB. In this scenario, the coefficient estimation is not biased by the CMB realization, and the fluctuations from the CMB component are removed so that the differences between methodologies are clearly shown. As expected, in terms of the resulting map, the \quilc\ approach provides the lower value of $\left\langle |\hat{P}_{\mathrm{CMB}}|^2 \right\rangle$, because its $2N_{\nu}$ degrees of freedom fit independently the $Q$ and $U$ combinations, in such a way that each coefficient estimation is affected individually by the particular realizations. On the other hand, the \pilc\ approach takes into account more physical constraints. Let us remark that the \pilc\ methodology provides the proper treatment of the residual contribution because the Stokes parameters are combined taking into account their spinorial properties. Notice that, as the foreground realization from the PSM does not show a global frequency-dependent shift of the polarization angle, the $\boldsymbol{\omega}^{(I)}$ coefficients take small values, such that the methodology is effectively weighting the polarization modulus with a half of degrees of freedom less than \quilc. Finally, of all the considered methodologies, the resulting maps from \prilc\ present the higher values of $\left\langle|\hat{P}_{\mathrm{CMB}}|^2\right\rangle$. Although systematically higher, they are similar to those obtained with \pilc. The improvement with \pilc\ is small because, in this situation, the two methodologies perform almost the same. For a specific set of simulations, the logarithm with base $10$ of the $\mathrm{\mathbf{C}}^{(+)}$ and $\mathrm{\mathbf{C}}^{(-)}$ matrices is shown in Figure~\ref{fig:matrix}. For the expected foreground characteristics, the first matrix is clearly dominant with respect to the second one, and therefore $\boldsymbol{\omega}^{(I)}$ are close to zero. The higher frequencies correspond to greater noise levels, and thereupon greater correlation.   

\subsubsection*{Angular power spectrum of the residuals}
\begin{figure}
\includegraphics[scale=0.67]{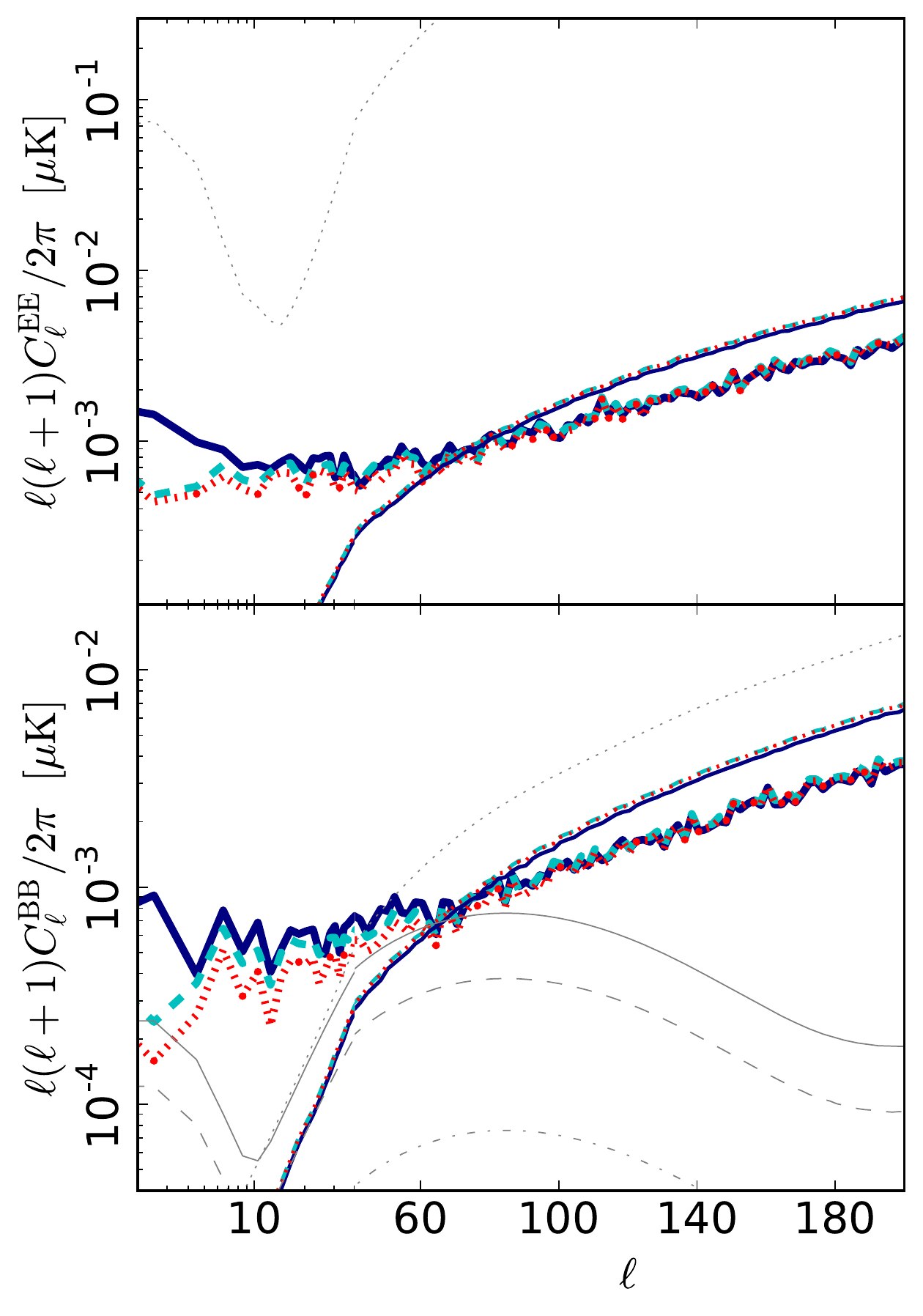}
\caption{Mean value of the EE (top panel) and BB (bottom panel) power spectra of the foreground (thick lines) and instrumental noise residuals (lines with intermediate thickness) from $100$ sets of simulations. The \quilc\ residuals are depicted by the solid lines (navy blue), the dashed lines (cyan) correspond to the \pilc\ residuals, and the dotted lines (red) represent the \prilc\ approach. Several fiducial CMB models are plotted in grey (thin lines) corresponding with different values of $r$: the scalar E-mode contribution (top panel) and the pure B-mode lensing contribution (bottom panel) are depicted by the dotted line, and the primordial B-mode contribution is shown for $r = 1\times 10^{-3}$ (dash-dotted line), $r = 5\times 10^{-3}$ (dashed line), and $r = 1\times 10^{-2}$ (solid line). The angular-multipole range is shown in logarithmic scale up to $\ell = 40$.}
\label{fig:foreressim}
\end{figure}

The expected angular power spectra of the foreground and the instrumental noise residuals of the foreground-reduced maps are shown in Figure~\ref{fig:foreressim}. They are computed as the mean value from simulations, where the coefficients of each methodology are computed for each realization. The pseudo-power spectra are corrected using the polarization MASTER estimator \citep[see][]{Kogut2003}. 

The residual level is similar for \pilc\ and \prilc, although this latter one seems to be lower at low angular multipoles. Notice that this is not in contradiction with the fact that the \prilc\ approach provides a CMB estimation with a higher value of $\left\langle |\hat{P}_{\mathrm{CMB}}|^2\right\rangle$, because when the residual component is isolated to compute its power spectrum, only the $\left\langle |\hat{P}_{\mathrm{residual}}|^2\right\rangle$ contribution to the total minimum variance is considered. However, the cross-correlations between the CMB signal and the residuals contributions are not negligible because, as the coefficients are estimated for each set of simulations, spurious correlations appear for the specific realization. This known bias \citep[see][]{Saha2008, Delabrouille2009, Chiang2009} contributes differently to \pilc\ and \prilc, as it depends on the number of degrees of freedom. The lower the instrumental noise level is, the more evident the effect of the bias will be at low angular multipoles, because the coefficient estimation is less affected by the noise-dominant multipoles. In a realistic case, the nominal sensitivity of the experiment is a limiting factor, but the data maps could be filtered, for instance, with a more aggressive window function. The ITF approach allows one to reach a better compromise than the one obtained by the ILC between the resolution of the resulting map and the instrumental-noise influence because the templates can be filtered without loss of data resolution. 

\begin{figure}
\includegraphics[scale=0.67]{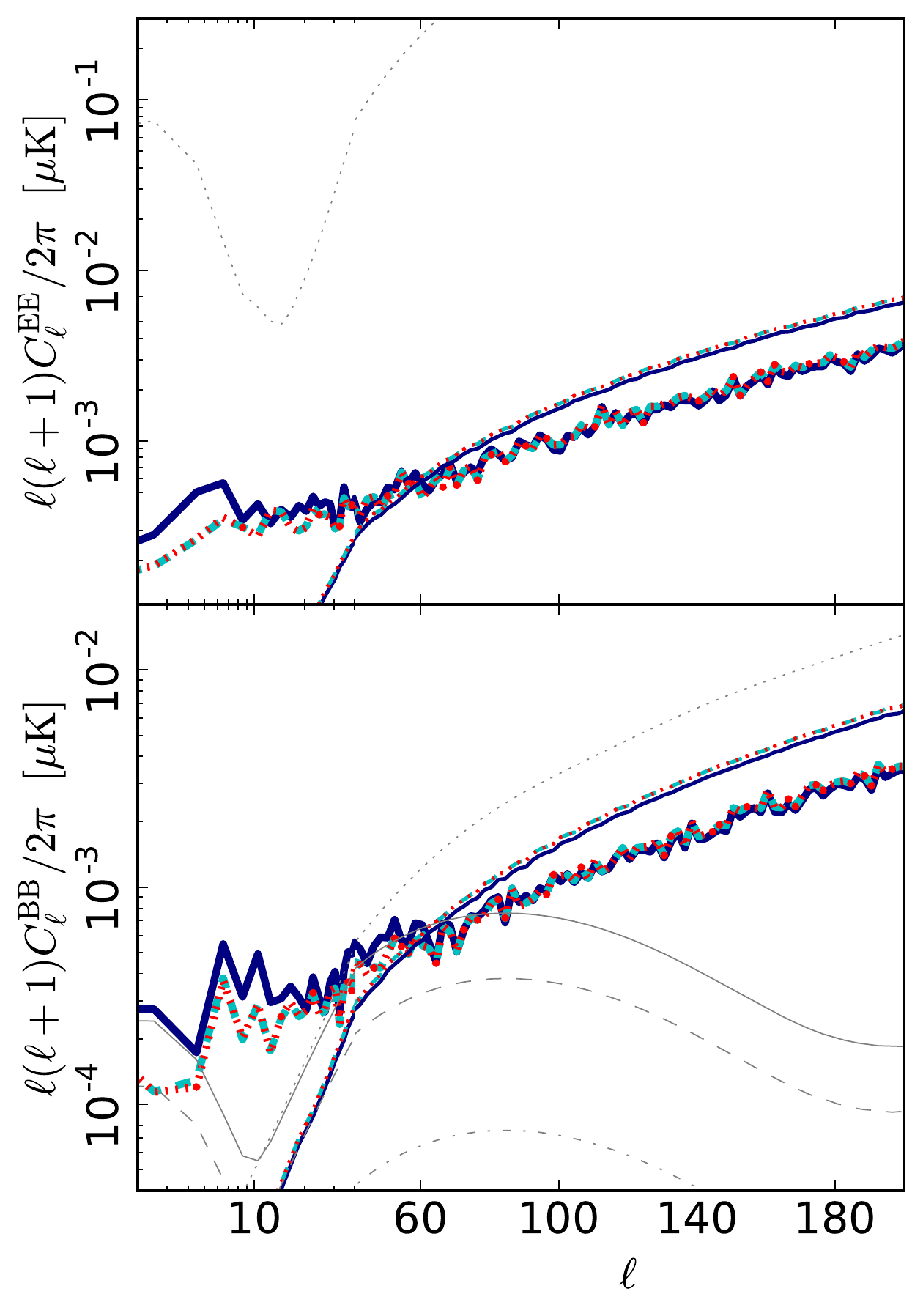}
\caption{Mean value of the EE (upper panel) and BB (bottom panel) power spectrum of the foreground (thick lines) and the instrumental noise residuals (lines with intermediate thickness) from $100$ sets of simulations in the ideal case in which the foreground and the instrumental noise covariances can be estimated (instead of considering the total map including the CMB component).  The \quilc\ residuals are depicted by the solid lines (in navy blue), the dashed lines (cyan) correspond to the \pilc\ residuals, and the \prilc\ residuals are represented by the dotted lines (in red). As in Figure~\ref{fig:foreressim}, several fiducial CMB models are plotted in grey (thin lines) corresponding with different values of $r$. The angular-multipole range is shown in logarithmic scale up to $\ell = 40$.}
\label{fig:foreressim_ilcfore}
\end{figure}

A way to avoid the most important contribution to this bias effect (i.e., the cross-correlation between the CMB realization and the foreground residual) is to explore an ideal case in which we could estimate the covariance matrix only taking into account the foreground and the instrumental noise contributions. In this case, only the cross-correlation between the foreground and the instrumental noise residuals contributes to the bias, but it can be considered negligible (at least in comparison with the one expected from the CMB and the foreground residual). The expected values of the residual angular power spectra obtained in this case are shown in Figure~\ref{fig:foreressim_ilcfore}. The total residual presents similar levels for all the methods. At low angular multipoles, \pilc\ and \prilc\ provide a lower power spectrum than \quilc, but this trend is inverted in the noise-dominant regime ($\ell \gtrsim 70$).

These plots show how important is a proper characterization of the foreground residual in the estimation of the tensor-to-scalar ratio $r$. The primordial B-mode polarization is depicted in the bottom panels by grey lines for different values of $r$. If we assume that we are able to characterize properly the residual level of our foreground-reduced maps, a likelihood can be used to quantify our uncertainty in the parameter estimation (in particular, in $r$ and in the foreground amplitude $A$). Considering a perfect estimation of the B-mode lensing and the instrumental noise biases, and a smoothed version of the mean value of our foreground residual contribution as foreground template, we obtain uncertainties of $\sigma (r) \sim 10^{-4}$, assuming a Gaussian likelihood for simplicity, which is valid at small scales. In practice, the error bar of the tensor-to-scalar ratio will be higher due to uncertainties in the foreground modelling and the delensing procedure \citep[see, e.g.][]{Errard2011}. As \pilc\ preserves the physical properties of the map contributions, this methodology could be useful in the estimation of the foreground residual.

\subsubsection*{Coefficients of the different ILC approaches}
\begin{figure*}
\includegraphics[scale=0.58]{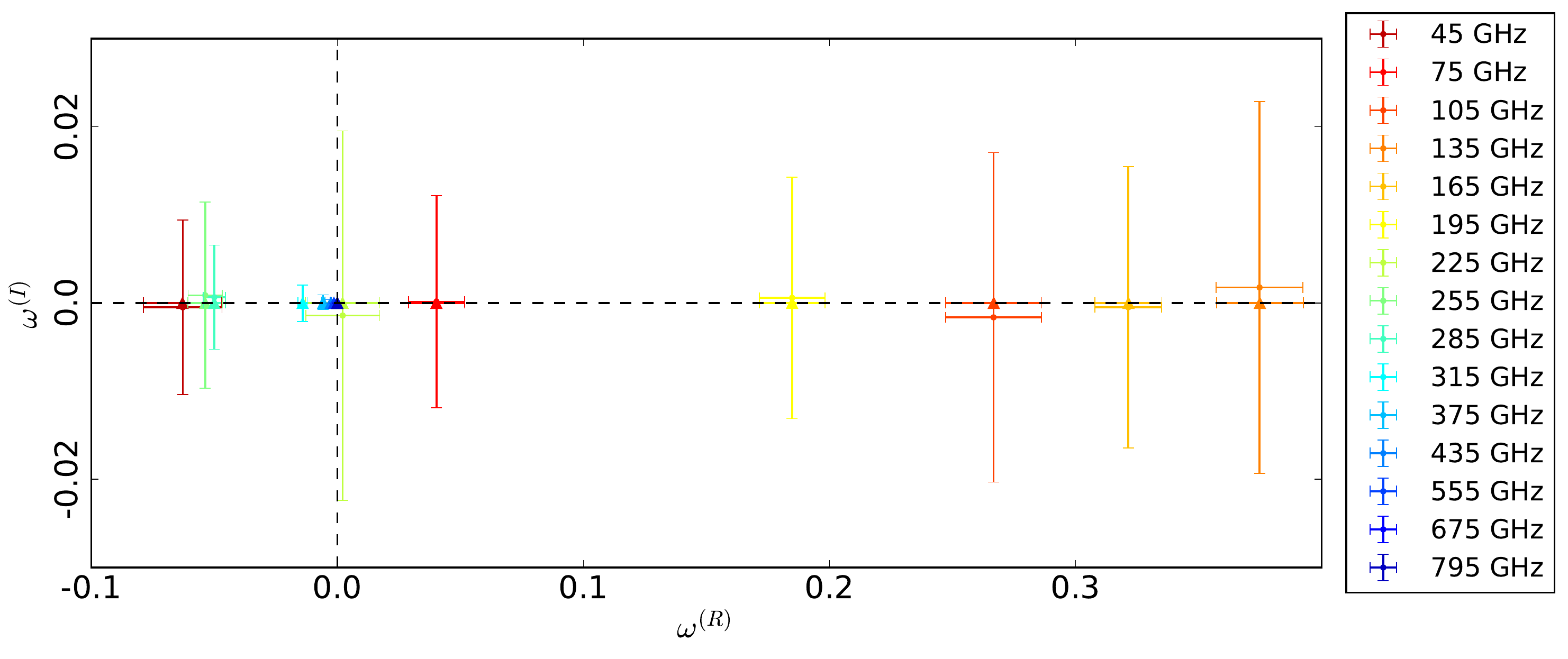} 
\caption{Mean values of the complex coefficients from simulations. The coefficients computed using \pilc\ are depicted by dots and those estimated with \prilc\ are plotted by triangles. The colour gradient represents the frequency range, from red (lower frequencies) to blue (higher frequencies). The error bars are estimated as the standard deviation from simulations. From left to right, the frequencies which are displayed are: $45\ \mathrm{GHz}$, $255\ \mathrm{GHz}$, $285\ \mathrm{GHz}$, $315\ \mathrm{GHz}$, $375\ \mathrm{GHz}$, $435\ \mathrm{GHz}$, $555\ \mathrm{GHz}$, $675\ \mathrm{GHz}$, $795\ \mathrm{GHz}$, $225\ \mathrm{GHz}$, $75\ \mathrm{GHz}$, $195\ \mathrm{GHz}$, $105\ \mathrm{GHz}$, $165\ \mathrm{GHz}$ and $135\ \mathrm{GHz}$.}
\label{fig:coef}
\end{figure*}

The study of the coefficients of the different ILC approaches provides further insight on the procedure. On the one hand, given a particular frequency, $\omega^{(Q)}$ and $\omega^{(U)}$ are of the same order for the \quilc\ approach. For the \pilc\ approach, in the case in which no frequency-dependent polarization rotation is present, the ensemble averages of the $\boldsymbol{\omega}^{(I)}$ coefficients are close to zero. However, focusing on a particular set of multifrequency simulations, the values of the coefficients are correlated with the CMB and the instrumental noise realizations, leading to the bias terms mentioned in the discussion about the power spectrum. These terms vanish when we take the ensemble average of the coefficients, since the mean values do not depend on the particular CMB realizations. In practice, these mean values should be the same in the realistic case (including the CMB signal in the covariance matrix) and in the ideal case in which the covariance matrix is estimated only with the foreground and the instrumental noise contributions. The mean values of the complex coefficients from simulated foregrounds from the PSM are shown in Figure~\ref{fig:coef} for the \pilc\ (depicted by dots) and \prilc\ (pure real coefficients, represented by triangles) methodologies. All coefficients are very close to the real axis. The corresponding ones to higher frequencies are lower because these channels present higher instrumental-noise contributions. The error bars are estimated as the standard deviation from simulations, and therefore they show the simulation-to-simulation variation of the coefficients due to the cosmic and the instrumental-noise variances. As the values of the coefficients which minimize the variance of $\hat{P}_{\mathrm{CMB}}$ simulation to simulation are deterministic, the fact that the error bars associated with the mean value of the imaginary parts of the coefficients are compatible with zero does not mean that this coefficients are dispensable. For a specific set of multifrequency realizations, the values of the $\boldsymbol{\omega}^{(I)}$ coefficients are typically lower than the values of the real parts $\boldsymbol{\omega}^{(R)}$, but non-zero. We discuss how their values increase when a global frequency-dependent phase on the foreground component is included in Section~\ref{subsec:toymodel}.

On the other hand, we test the case in which the \quilc\ approach is equivalent to minimize the expected value of $|\hat{P}_{\mathrm{CMB}}|^2$, as discussed in Section~\ref{sec:dis}. We have shown that, within the PSM, it is not expected a significant global deviation of the polarization phase depending on the frequency. The behaviour of the methodologies in this situation is also checked independently of the foreground emission. For this proposal, we use simulations with only CMB signal and a white noise contribution. In this case, the mean value of $\omega^{(I)}_j$ for each frequency $\nu_j$ is null for the \pilc\ method. For the \quilc\ approach, the expected values of $\omega^{(Q)}_j$ and $\omega^{(U)}_j$ are equal to each other and the same as $\omega^{(R)}_j$. However, let us remind that, even in this case, minimizing separately $\left\langle \hat{Q}^2_{\mathrm{CMB}} \right\rangle$ and $\left\langle \hat{U}^2_{\mathrm{CMB}} \right\rangle$ is not equivalent to minimize $\left\langle |\hat{P}_{\mathrm{CMB}}|^2 \right\rangle$, because non-invariant terms are considered in the first case.

\subsection{Toy model of polarization rotation}
\label{subsec:toymodel}
Finally, to show the potential of the \pilc\ methodology and, in particular, what can be learnt from the imaginary part of the coefficients, we use a toy model which presents a global shift on the polarization angle with a frequency dependence proportional to $\nu^{-2}$, motivated by the Faraday rotation \citep[see, e.g.][]{Oppermann2015}. In particular, we rotate the polarization of the PSM foregrounds a global phase shift with the following parametrization: 
\begin{equation}
\phi(\nu) = R_{\mathrm{45}}\left(\dfrac{45\ \mathrm{[GHz]}}{\nu}\right)^2\ \mathrm{[deg]},
\end{equation} where $R_{\mathrm{45}}$ is the angle in degrees of the polarization rotation corresponding to the $45\ \mathrm{GHz}$ channel. If we compare this expression with the one that describes the phase shift induced by the Faraday rotation, we find that, within this simile, $R_{\mathrm{45}}$ is proportional to the rotation measure $\mathrm{RM}$. However, it is necessary to note that this toy model is only intended to show roughly the properties of the \pilc\ proposal, and it does not actually correspond to a real Faraday rotation. First, the rotation considered here is a global polarization phase, which, in the context of the Faraday rotation, would be identified with a uniform magnetic field. Secondly, the Faraday rotation affects only to specific foreground components, such as the synchrotron, whilst in this model all the foreground components are rotated. And finally, the CMB component remains unrotated. In a realistic case, the CMB is also affected by the Faraday rotation, and an unbiased recovery requires additional considerations, such as taking into account the proper frequency dependence of the component to be recovered in the constraints.

\begin{figure}
\includegraphics[scale=0.7]{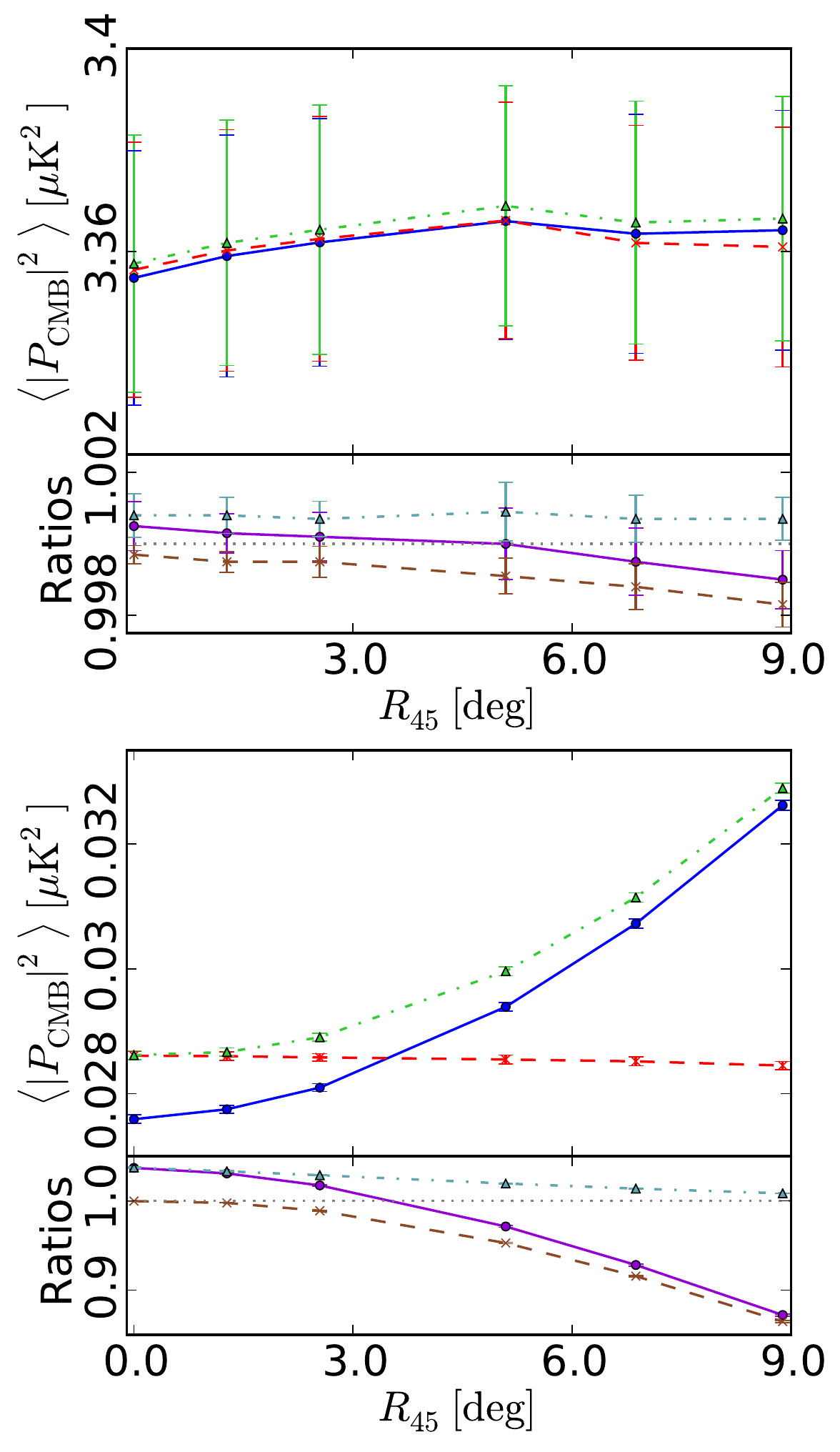}
\caption{Mean values of the variance of $\hat{P}_{\mathrm{CMB}}$ of the resulting maps as a function of $R_{\mathrm{45}}$ from \quilc\ (solid blue lines), \pilc\ (dashed red lines) and \prilc\ (dash-dotted green lines). The ratios between the mean values of the variances from the different methods are plotted in the smaller bottom panels, where the purple line (solid line) represents the ratio between \pilc\ and \quilc\, the brown line (dashed line) depicts the ratio between \pilc\ and \prilc\, and the fountain blue line (dash-dotted line) plots the ratio between \quilc\ and \prilc. The ideal case in which the covariance matrix can be estimated only with the instrumental noise and the foreground components is shown in the bottom panels.}
\label{fig:var_RM}
\end{figure}

In Figure~\ref{fig:var_RM}, the mean value of the variance of $\hat{P}_{\mathrm{CMB}}$ is shown as a function of the $R_{\mathrm{45}}$. As mentioned in the previous section, the intrinsic fluctuations of the variance are greater than the differences between methods. The cosmic variance and the instrumental noise uncertainties are represented in the error bars computed as the standard deviation from simulations. The mean value of the ratios between the variances from the different methods are plotted in the smaller bottom panel. As the differences are not evident due to the cosmic variance, the ideal case in which the covariance matrix can be estimated only with the instrumental noise and the foreground components is shown in the lower panel, where the standard deviations account only from the instrumental noise uncertainties. The greater the $R_{\mathrm{45}}$ value is considered, the greater the variance of the resulting map from \quilc\ and \prilc\ is, whilst the variance of the resulting map from \pilc\ remains constant. The \pilc\ methodology provides a better solution in terms of the variance from values of $R_{\mathrm{45}}\approx 4.0\ \mathrm{deg}$ (which corresponds to $\mathrm{RM} \approx 1500\ \mathrm{m^{-2}}$ in the rough simile with the Faraday rotation). For extreme cases with $R_{\mathrm{45}}\approx 7.5\ \mathrm{deg}$, the differences between the variances of the methodologies are even visible in spite of the cosmic variance from the CMB signal, and they are also present in terms of the power spectrum of the residuals.

\begin{figure*}
\includegraphics[scale=0.58]{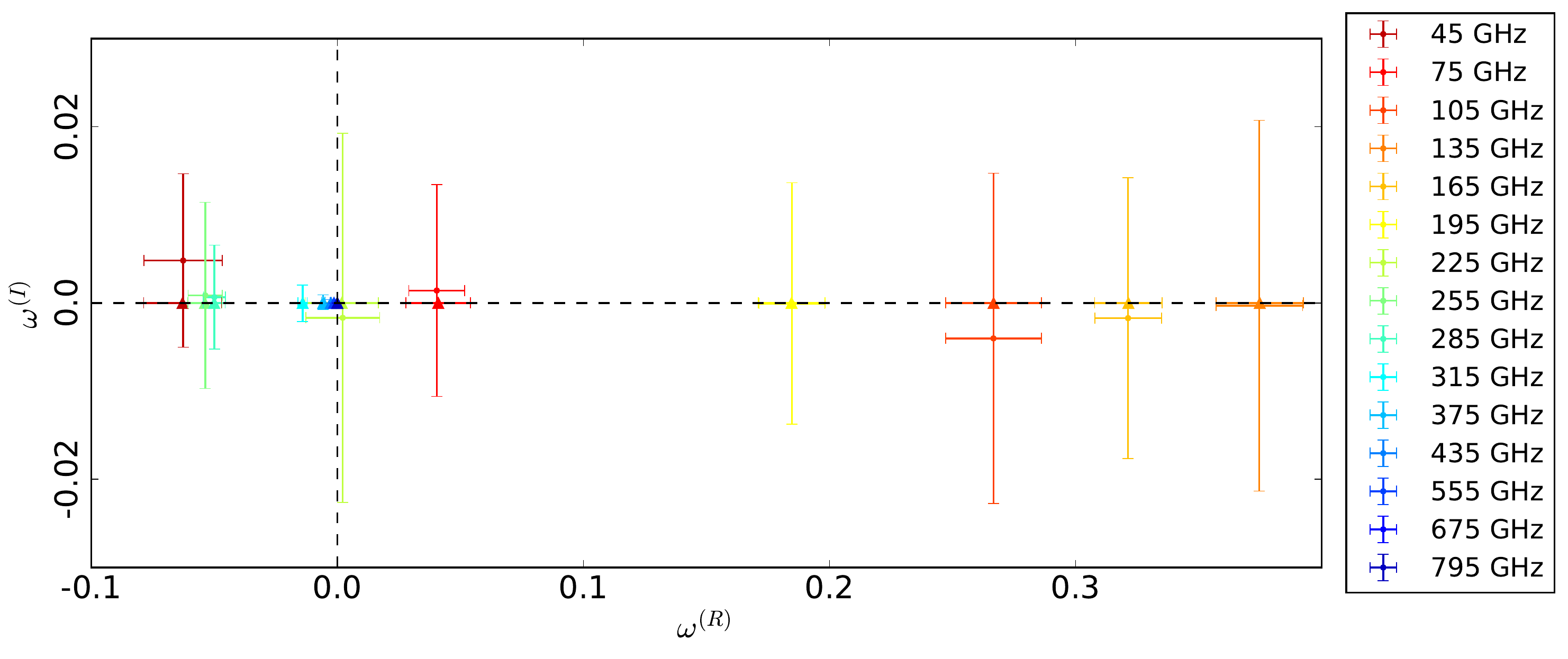} 
\caption{Mean values of the complex coefficients from simulations with a polarization rotation in the foreground component following a toy model which reproduces a global polarization rotation with $R_{\mathrm{45}}= 2.55\ \mathrm{deg}$. The coefficients computed using \pilc\ are depicted by dots and those estimated with \prilc\ are plotted by triangles. The colour gradient represents the frequency range, from red (lower frequencies) to blue (higher frequencies). The error bars are estimated as the standard deviation from simulations. From left to right, the frequencies which are displayed are: $45\ \mathrm{GHz}$, $255\ \mathrm{GHz}$, $285\ \mathrm{GHz}$, $315\ \mathrm{GHz}$, $375\ \mathrm{GHz}$, $435\ \mathrm{GHz}$, $555\ \mathrm{GHz}$, $675\ \mathrm{GHz}$, $795\ \mathrm{GHz}$, $225\ \mathrm{GHz}$, $75\ \mathrm{GHz}$, $195\ \mathrm{GHz}$, $105\ \mathrm{GHz}$, $165\ \mathrm{GHz}$ and $135\ \mathrm{GHz}$.}
\label{fig:coef_RM1000}
\end{figure*}

In the following lines, the particular case with a polarization rotation with $R_{\mathrm{45}}= 2.55\ \mathrm{deg}$ is explored in terms of the coefficients of \pilc\ (this corresponds to a $\mathrm{RM} = 1000\ \mathrm{m^{-2}}$, value that one could expect to reach in a realistic case from the Faraday rotation of the synchrotron component in specific regions of the Galactic plane). The complex coefficients corresponding to this case are shown in Figure~\ref{fig:coef_RM1000}. The imaginary parts of these coefficients become more significant than the values obtained from the pure PSM foreground contribution plotted in Figure~\ref{fig:coef}. The greatest deviation from the real axis corresponds to the lower frequencies, which suffer a greater polarization rotation. For this value of $R_{\mathrm{45}}$, all the standard deviations of the imaginary part are still compatible with zero. Whilst these deviations depend on the cosmological model, the greater the $R_{\mathrm{45}}$ value is considered, the greater the imaginary parts of the coefficients are, in such a way that, for a value of $R_{\mathrm{45}} = 5.0\ \mathrm{deg}$, the corresponding coefficient of the $45 \ \mathrm{GHz}$ channel is deviated from the real axis a distance in terms of the standard deviations greater than the $1\sigma$ level. As said in the previous section, these deviations are due to the differences between the realization-to-realization coefficient estimation. For a particular set of realizations, these coefficients are deterministic, and their imaginary parts are greater when the $R_{\mathrm{45}}$ increases. We also observe that, in general, the modulus of the complex coefficient is preserved as $R_{\mathrm{45}}$ grows.

Summarizing, the \pilc\ methodology provides a better solution in terms of the variance of $P$ in the resulting maps when a global shift is considered in the polarization angle. When we make the comparison with the \quilc\ and \prilc\ approaches, the rotation effects on the foreground-reduced maps are more visible as the shift increases. In terms of the coefficients, the imaginary parts become more important. In the case in which the frequency range is extended to lower values (as those which are necessary to monitor the synchrotron emission) the effect of the rotation should be more significant. However, the cosmic variance, as well as the constraint imposed to the coefficients of the linear combination, may hamper a proper estimation of the rotation angle without a specific optimization of the method to recover this component instead of the CMB signal. 

\section[Conclusions]{Conclusions}
\label{sec:Conclusions}
We present a linear combination approach in which the two-spin quantity $Q\pm iU$ is combined with complex coefficients to obtain a recovery of the CMB signal from a set of frequency maps. Although, in this paper, this scheme is only considered for two linear-combination approaches (the ILC and the ITF), working on the $P$ map instead of using $Q$ and $U$ separately could be applied to other component separation methodologies, both to those based on linear combinations such as \smica, and all those which do not, such as, for instance, methods based on neural networks or parametric approaches (in the sense that the foreground polarization models should be covariant). It works directly on the $Q$ and $U$ Stokes parameter maps, enabling to deal with data from a partial sky-coverage without resorting to the harmonic space. The coefficients are computed by minimizing the expected value of $\hat{P}^2_{\mathrm{CMB}}$ in the resulting map. All the terms involved in the minimization are covariant quantities, in contrast to those terms which appear when the expected values of $\hat{Q}^2_{\mathrm{CMB}}$ and $\hat{U}^2_{\mathrm{CMB}}$ are separately minimized. 
 
In forthcoming CMB polarization experiments, the residual level of foregrounds will depend on the particular properties of each experiment, such as its sensitivity and resolution, the sky coverage, the frequency range or the number of channels. For some of these configurations, as the residual component could be at the level of the CMB signal, it might be necessary to model the foreground residuals to statistically remove its contribution from the CMB estimation. The new ILC methodology preserves the coherence between the two spinorial components, such that the physical meaning of the residual is guaranteed. On the contrary, removing foregrounds independently in $Q$ and $U$ requires multiplying the Stokes parameters by different coefficients. As they are quantities which depend on the local coordinate frame, this implies to change arbitrarily the polarization angle and modulus, spoiling the physical description of the residual polarization.

Within the \pilc\ approach, in the standard case in which there is not a dominant component with a global frequency-dependent phase in the polarization angle, the set of $N_{\nu}$ coefficients associated with the real parts of the complex coefficients, $\boldsymbol{\omega}^{(R)}$, are weighting the polarization modulus, and they are directly comparable with the coefficients computed using the \quilc\ approach (although, in the first case, $\left\langle Q_{\mathrm{CMB}}^2 + U_{\mathrm{CMB}}^2 \right\rangle$ is minimized, whilst $\left\langle Q_{\mathrm{CMB}}^2\right\rangle$ and $\left\langle U_{\mathrm{CMB}}^2\right\rangle$ are separately considered in the second approach). The imaginary parts of the coefficients allow E-B mixing depending on the frequency, and they arise from considering a non-vanishing $\left\langle Q_iU_j-Q_jU_i\right\rangle$ cross-correlation. Their values become greater when a different global polarization rotation is applied at each frequency band. Therefore, this two-spin methodology could be useful to remove and estimate the contribution of the Faraday rotation in particular regions with coherent magnetic field.

The methodology is tested on sets of multifrequency simulations. In terms of the power spectrum, the residual levels obtained from both the new proposed method and the standard implementation of the ILC are similar. As no dominant global phase shift with a frequency dependence is present in the foreground components simulated with the PSM, we also test the \prilc\ approach, in which the $\boldsymbol{\omega}^{(I)}$ coefficients are set to zero. In this situation, the \prilc\ approach is equivalent to minimize jointly the expected value of $\hat{Q}^2_{\mathrm{CMB}}+\hat{U}^2_{\mathrm{CMB}}$ with the same coefficients associated with both Stokes parameter maps.

Finally, a toy model of a global polarization rotation is considered to show the potential of the \pilc\ methodology. In terms of the variance of the resulting maps, \pilc\ provides clearly lower contributions than \quilc\ and \prilc\ for values of the phase shift in the lowest frequency of $R_{\mathrm{45}}\approx 4.0\ \mathrm{deg}$. However, the methodology will be optimized in a future work to detect the effect of a frequency-dependent polarization rotation in more realistic scenarios.
  


\section*{acknowledgments}
The authors thank Marcos L\'opez-Caniego for his assistance with foreground simulations, and Beatriz Ruiz-Granados for useful discussions. Partial financial support from the Spanish Ministerio de Econom\'{i}a y Competitividad Projects AYA2010-21766-C03-01, AYA2012-39475-C02-01 and Consolider-Ingenio 2010 CSD2010-00064 is acknowledged.
\bibliographystyle{mn2e}
\bibliography{citas_ilc_pol}

\end{document}